# Nanoscale Phase Distribution Governs Exchange Bias in Multiphase Magnetic Nanoparticles

Murtaza Bohra[1], Stefanos Giaremis[2,3], Vidyadhar Singh[4], Joseph Kioseoglou[2,3], Stephan Steinhauer[5] and Panagiotis Grammatikopoulos[6*]

[1]*Ecole Centrale School of Engineering, Mahindra University, Survey Number 62/1A, Bahadurpally Jeedimetla, Hyderabad 500043, Telangana, India*

[2]*School of Physics, Department of Condensed Matter and Materials Physics, Aristotle University of Thessaloniki, 54124 Thessaloniki, Greece*

[3]*Center for Interdisciplinary Research and Innovation, Aristotle University of Thessaloniki, 57001 Thessaloniki, Greece*

[4]*Department of Physics, Jai Prakash University, Chapra, 841301 Bihar, India*

[5]*Department of Applied Physics, KTH Royal Institute of Technology, Albanova University Centre, Roslagstullsbacken 21, 106 91 Stockholm, Sweden*

[6]*Instituto Regional de Investigación Científica Aplicada (IRICA), Departamento de Física, Universidad de Castilla-La Mancha, 13071 Ciudad Real, Spain*

E-mail: p.grammatikopoulos@uclm.es

**ABSTRACT:** Exchange bias at ferromagnet–antiferromagnet interfaces underpins magnetic memory, spintronic devices, and nanoscale electromagnetic technologies, yet its behaviour in complex nanoscale heterostructures remains poorly understood. Here we uncover how exchange bias emerges in functional multiphase metal–oxide nanoparticles by combining gas-phase synthesis, advanced magnetic characterisation, and first-principles-informed spin-dynamics simulations. Using Ni–Cr/NiO nanoparticles as a model system, we show that exchange bias is governed not simply by the presence of ferromagnetic and antiferromagnetic phases, but critically by their nanoscale spatial distribution and interfacial topology. At 10 K, significant negative exchange bias (0.8 kOe) and coercivity enhancement (1.4 kOe) was exhibited; both decreased due to either Cr-segregation (at low Cr content) or to Cr accumulation inside the core (at high Cr content). The resulting competition between magnetic phases produces a temperature-driven inversion from negative to positive exchange bias and a crossover from exchange-dominated to dipolar interactions. By linking density-functional-theory calculations directly to spin-dynamics simulations of nanoparticle ensembles, we establish a predictive framework for designing exchange-coupled nanomagnets capable of operating beyond the superparamagnetic limit.

## 1 | Introduction

Magnetic exchange bias, resulting in increased coercivity and hysteresis loop shifts along the magnetic field axis, typically occurs at the interface between a ferromagnet and an antiferromagnet [1]. Adequate control of interface exchange bias couplings poses a significant challenge for ground-breaking new applications. To this end, in-depth understanding of magnetic ordering states is decisive, whether the structures are ferromagnetic (FM), antiferromagnetic (AFM), superparamagnetic (SPM), or spin glass (SG) at room temperature.

Presently, with a growing demand for 6G-7G technologies, the Internet of Things (IoT), ultrahigh-density memory, and eco-friendly magnets, reliance on exchange-coupled core-shell nanoparticles (C/S-NPs) is expected to rise [2-6]. Traditionally, C/S-NPs feature a FM core and an AFM shell, but the reverse magnetic configuration is also possible. However, the existence of multiple magnetic phases in the core and shell regions may lead to different interface interactions, adding complexity but, at the same time, offering tuning capabilities for the exchange bias.

As a result, current research is exploring not only various materials and material combinations for C/S-NPs but also diverse approaches to maximise their exchange bias effect. Such efforts in Co/CoO C/S-NPs were made by embedding them into NiO [7] and $Cu_2O$ [8] matrices. In the former case, the high Néel temperature ($T_N$) of the NiO matrix induced a magnetic proximity effect, resulting in an effective AFM character with an apparent $T_N$ exceeding that of bulk CoO. In the latter case, lattice matching between $Cu_2O$ and CoO played a critical role in two aspects: structurally stabilising the CoO shell and promoting a significant presence of uncompensated moments within the CoO shell. On the contrary, reverse Ni/NiO/CoO C/S-NP systems revealed a magnetic proximity effect between NiO and CoO, which became more pronounced as the thickness of the intermediate NiO shell increased [9]. Lately, positive exchange bias was observed in the $Gd_xFe_{1-x}$/NiCoO FM/AFM system at low cooling fields due to oxygen ion migration from NiCoO to the Gd [10]. This concept was also applied to achieve magnetoionic control of perpendicular exchange bias in a NiO/Pd/Co/Pd/$Gd(OH)_3$/Au system using voltage-mediated hydrogen ion pumping [11]. Recent gas-phase synthesis of M(Fe, Co, and Ni)Cr NPs [12] has unveiled intriguing phenomena, such as stabilising the high-temperature FeCr phase as a seed in watermelon-type C/S-NPs structures [13], forming complete Cr/Co C/S-NPs in CoCr NPs with a surface-to-volume ratio < 1 [14], and Cr satellites encircling NiCr NPs [15]. These results imply that Cr-atom segregation toward the surface leads to diverse core/shell structures. However, through additional doping with 5 at. % Cr, it is possible to retrieve bulk NiCr characteristics with Cr distributed in the Ni core rather than forming Cr satellites [16].

Inspired by these studies, here we investigate the magnetic exchange bias in gas-phase synthesised Ni-Cr/NiO C/S-NPs with a tuneable Curie temperature ($T_C$) by varying the Cr doping in the FM core. In parallel, we derive from first-principles calculations fundamental magnetic properties of all observed phases and use them to inform spin-dynamics models corresponding to the experimental C/S-NP ensembles; namely, Ni and/or Ni-Cr as the FM cores and NiO ($T_N$ = 525 K) embedded with Cr ($T_N$ = 230 K), and $Cr_2O_3$ ($T_N$ = 307 K) as the AFM shells. Our facile methodology (available as **Supporting Information**) allows for the construction of flexible models for combined experimental and theoretical studies of the exchange bias effect. The framework introduced here provides a route for designing exchange-coupled nanomagnets capable of operating beyond the superparamagnetic limit and offers a general strategy for connecting first-principles electronic structure calculations with the magnetic behaviour of realistic nanoparticle ensembles.

## 2 | Predictive framework for exchange bias in complex C/S-NPs

The central concept and methodological novelty of this work are schematically illustrated in **Figure 1**. Two parallel pipelines, one experimental and one computational, are signified by the green and yellow arrows, respectively. Multiphase C/S-NPs are initially deposited by magnetron-sputtering inert-gas condensation (step 1), yielding tailored nanoparticulate samples of certain variability in size and configuration (step 2) with corresponding magnetic properties (step 3). After meticulous characterisation, all present phases in the sample are identified as schematically summarised (step 4), and their magnetic properties are computed by first-principles calculations. Statistical-weight averaging informed by experimental data result in approximate digital analogues of the deposited samples (step 5). Spin dynamics simulations of these digital analogues produce ensemble magnetisation loops (step 7); the computational workflow is summarised in step 6. Feedback from our simulated ensemble magnetisation loops may inform subsequent depositions (step 8), exploiting the high-level control available with the sophisticated gas-phase synthesis method. Our facile predictive framework indicates that the spatial distribution of Cr determines the strength and sign of exchange bias (step 9).

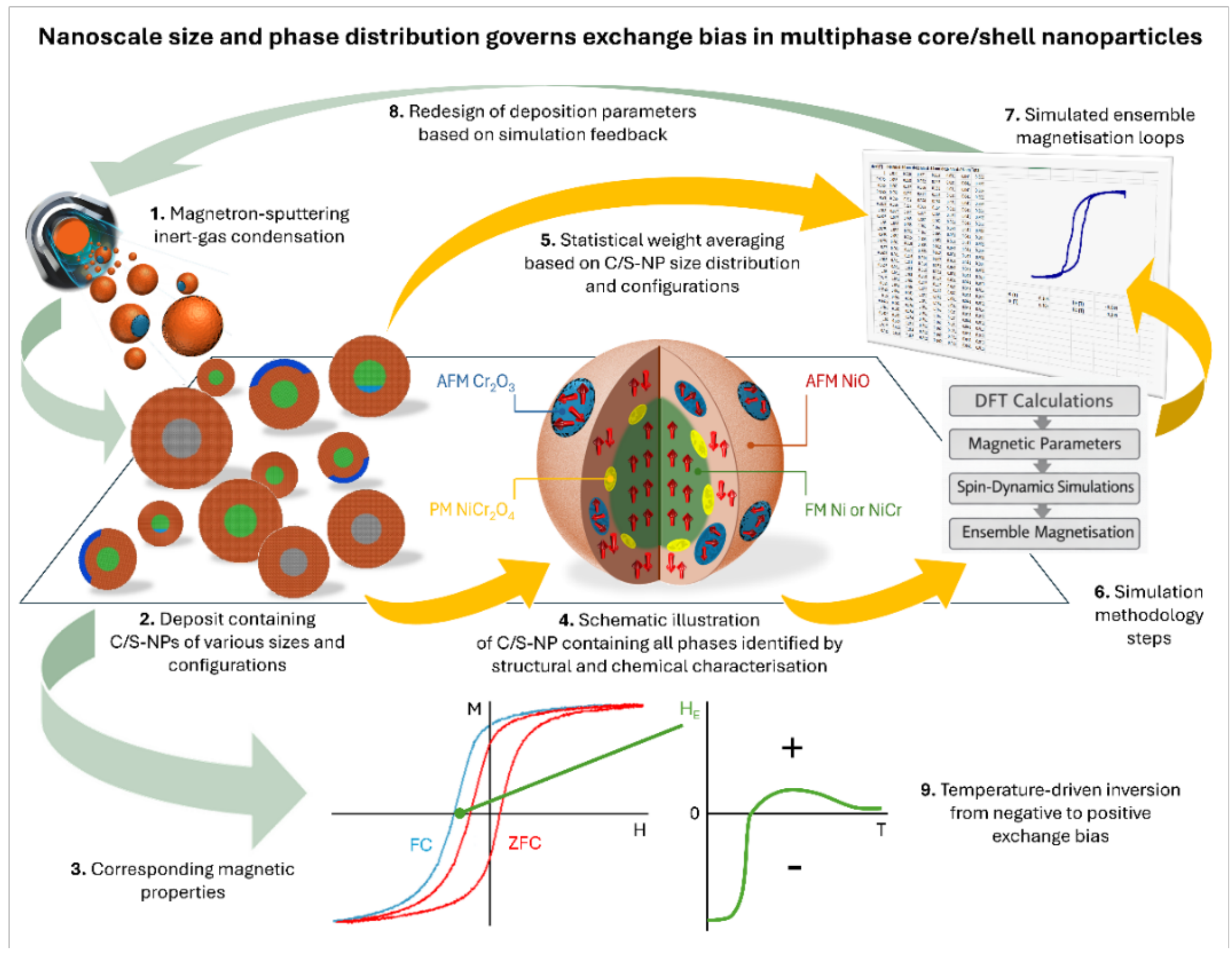

**Figure 1 | Nanoscale phase distribution controls exchange bias in multiphase nanoparticles.** Experimental and computational workflows are indicated by green and yellow arrows, respectively. Gas-phase synthesis and deposition produces samples of multiphase nanoparticles with corresponding magnetic properties. Detailed characterisation and statistical analysis of the samples inform their digital analogues, the magnetic behaviour of which is calculated with a combination of first-principle and spin dynamics simulations. Simulated samples may provide feedback for subsequent depositions of new CS-NP samples with tailored exchange bias. Step 4 presents a schematic illustration of C/S-NP, containing all phases identified by structural and chemical characterisation.

## 3 | Structure and chemical composition of C/S-NPs

The structure and surface chemical composition of the C/S-NPs was investigated in detail using X-ray-based techniques. **Figure 2a** shows grazing incidence X-ray diffraction (GIXRD) patterns of $Ni_{1-x}Cr_x$ ($0 \geq x \geq 0.15$) C/S-NPs, juxtaposed with International Centre for Diffraction Data (ICDD) records for bulk Ni and NiO for comparison [16]. Three broad peaks can be identified; two peaks around 2θ of 37.06° and 62.62° may be indexed to NiO and an overlap of 43.09° and 44.53° peaks may belong to Ni and NiO phases, respectively. However, matching of few XRD peaks with other $Cr_2O_3$ and $NiCr_2O_4$ phases is not clearly discernible by XRD alone. The lattice constants calculated from the 43.09° peak width fall between 4.13–4.16 Å, against bulk NiO (4.2 Å) and Ni (3.5 Å).

X-ray photoelectron spectroscopy (XPS) core level spectra of $Ni_{2p}$, $Cr_{2p}$ and $O_{1s}$ appear asymmetric, as shown in Figure 2b-d, with multiple oxide phases. To quantify in terms of the Ni/Cr atomic ratios in metallic alloys and oxides, XPS data were deconvoluted (Supporting Information **Figure S1**) and the following observations were made: (i) the Ni $2p_{3/2}$ peak, located at binding energy around 852.9 eV, corresponds to pure Ni metal, while 856.39 eV and 861.92 eV peaks correspond to NiO and NiO-satellite, respectively. (ii) The broad Cr $2p_{3/2}$ peak composed of three energies (577.2, 576, and 574.2 eV) contains contributions from $Cr^{3+}$, $Cr^{2+}$, and $Cr^{0}$, respectively [16]. This indicates formation of $Cr_2O_3$ and $Cr_2O$ along with metallic Cr. (iii) The $O_{1s}$ core level demonstrates three components. The low binding energy component refers to the lattice oxygen atoms, whereas the two high binding energy components correspond to the surface adsorbed C-O groups resulting from moisture and air annealing process. To estimate the composition of the interior core further, the C/S-NPs were *in situ* etched in the XPS chamber under high vacuum and XPS spectra were measured again (Figure S1). Cr was then detected at the cores and metallic (NiCr) phases increased after etching the surface. $NiCr_2O_4$ was also detected after the etching process. Thus, XPS results suggest that metallic Ni-Cr core is surrounded by a mixture of Ni and Cr oxides shell (NiO and $Cr_2O_3$), along with $NiCr_2O_4$ and suboxides (metastable phases).

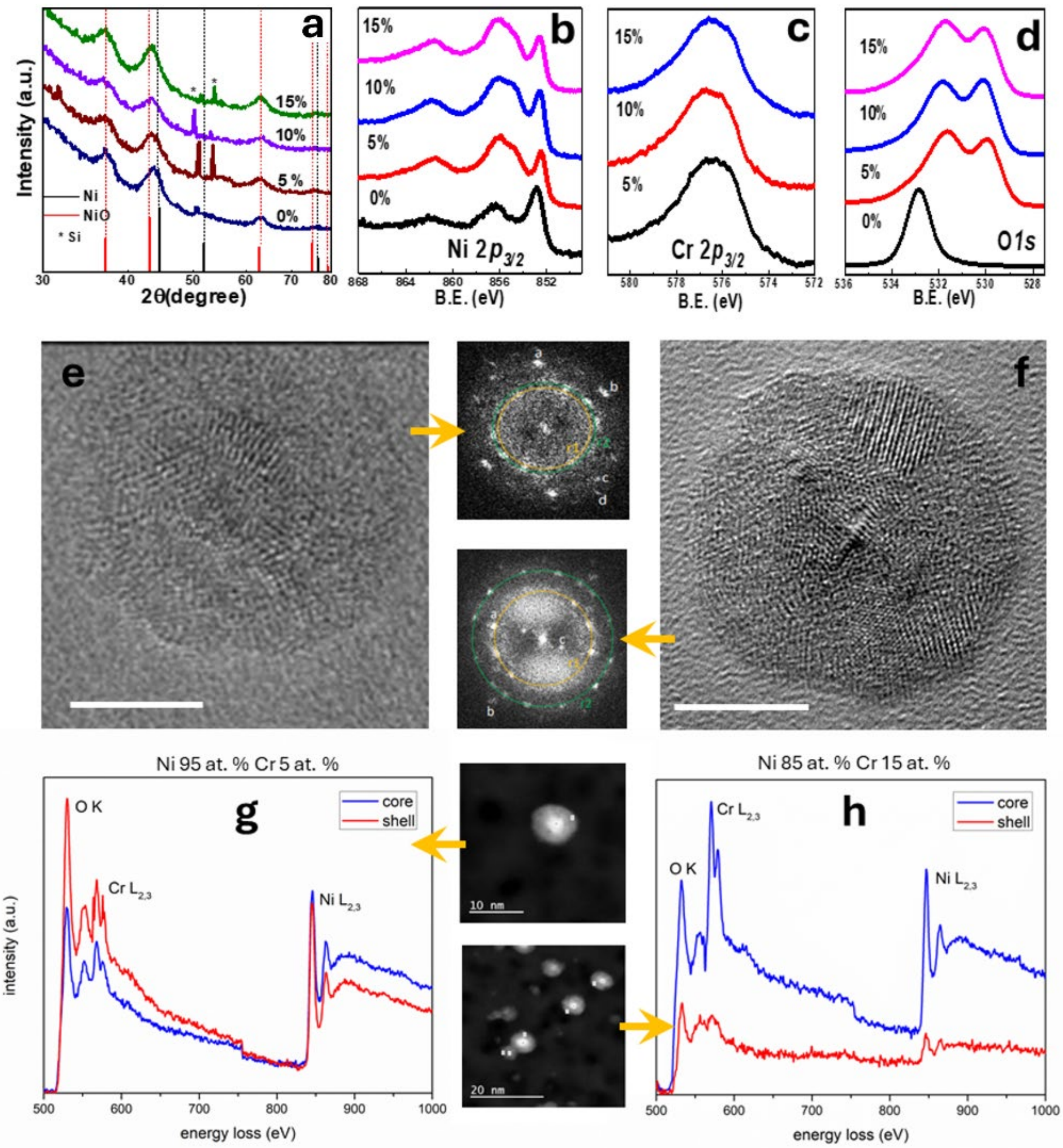

**Figure 2 | Structure and chemical composition of C/S-NPs. a)** GIXRD of $Ni_{1-x}Cr_x$ ($0 \geq x \geq 0.15$) C/S-NPs along with ICDD reference data for Ni and NiO for comparison. **b-d)** XPS spectra showing the peaks of Ni 2p, Cr 2p and O1s for $Ni_{1-x}Cr_x$ ($0 \geq x \geq 0.15$) C/S-NPs. **e-f)** HRTEM images and corresponding FFT analysis of two representative $Ni_{1-x}Cr_x$ samples (x=0.05 and 0.15). Scale bars: 5 nm. **g-h)** HAADF-STEM images and corresponding EELS analysis confirming core-shell structures, with cores composed of metallic Ni and NiCr and shells composed of oxides of Ni, Cr and Ni-Cr-O.

High-resolution and high-angle annular dark-field scanning transmission electron microscopy (HRTEM and HAADF-STEM, respectively) images (Figure 2e-h) of two representative samples show core-shell nanostructures with typical core and shell sizes of ~5 nm and ~3 nm, respectively. More exemplar TEM images are shown in Supporting Information **Figure S2**. Fast Fourier Transform (FFT) analysis of the HRTEM images indicate the polycrystalline nature of core and shell, composed of multiple oxide materials (Supporting Information

Table S1). Elemental mappings (via electron energy loss spectroscopy, EELS) are consistent with the NP core-shell structure and the aforementioned Ni and Cr metallic / oxide phases. Chromium is present across the core and shell of the NiCr-5% sample, while the core of the NiCr-15% sample contains more Cr. This result agrees with reported metallic NiCr NPs [16-17], where Cr was more segregated in Cr-5% than in Cr-15% samples. However, it also indicates that even air annealing at 200 ºC was not sufficient to extract all Cr from the core because of the screening effect.

In conclusion, based on XRD, XPS, and TEM characterisation studies, we schematically illustrated our C/S-NPs (Figure 1), which consist of a metallic FM core of Ni or Ni-Cr surrounded by an AFM NiO shell containing embedded AFM $Cr_2O_3$ and interfacial paramagnetic (PM) $NiCr_2O_4$.

## 4 | Magnetic exchange bias in C/S-NPs

Exchange bias coupling normally produces an additional unidirectional anisotropy in the FM core below the $T_N$ of AFM shells and $T_C$ of FM cores.[2] To establish the presence of exchange bias, the C/S-NPs were heated up to 400 K and then cooled to 10 K in the presence of a 10 kOe fixed field. Field cooled (FC) *M-H* loops taken at 10 K are shown in **Figure 3a** (left; in the right, corresponding simulated loops are shown, as discussed in a following section). These *M–H* loops are shifted in the negative field direction due to the exchange interaction between the FM core and the AFM oxide shell. This effect is drastically reduced in Cr 15 at. % samples with a decreased $H_C$ value, indicating that Cr may be fully present in the FM core, and approaching the PM bulk NiCr behaviour. In such a FM/AFM exchange system, the AFM shell affects the shape of FC loops both at low–field through exchange bias and at higher–field with non-saturation effects. Noticeably, pure Ni-NiO C/S-NPs display "*hummingbird*"-type hysteresis consisting of shifted and unshifted *M–H* loops, indicating the presence of a mixture of soft magnetic (Ni) and hard magnetic (Ni–NiO) phases [18]. This stems from the fact that some of the Ni NPs were partially or hermetically protected from surface oxidation, possibly due to being surrounded by other Ni NPs; they are referred to as soft–magnetic Ni NPs. In contrast, other NPs were completely surface–oxidised and behaved more like hard–magnetic exchange–coupled Ni–NiO C/S-NPs [19]. However, this feature is not as markedly observed in Cr–doped C/S-NPs, because Cr segregation facilitated complete surface oxidation surrounding the FM metallic core. The large non-saturation in higher Cr–doping (15 at. %) C/S-NPs is ascribed to the PM state.

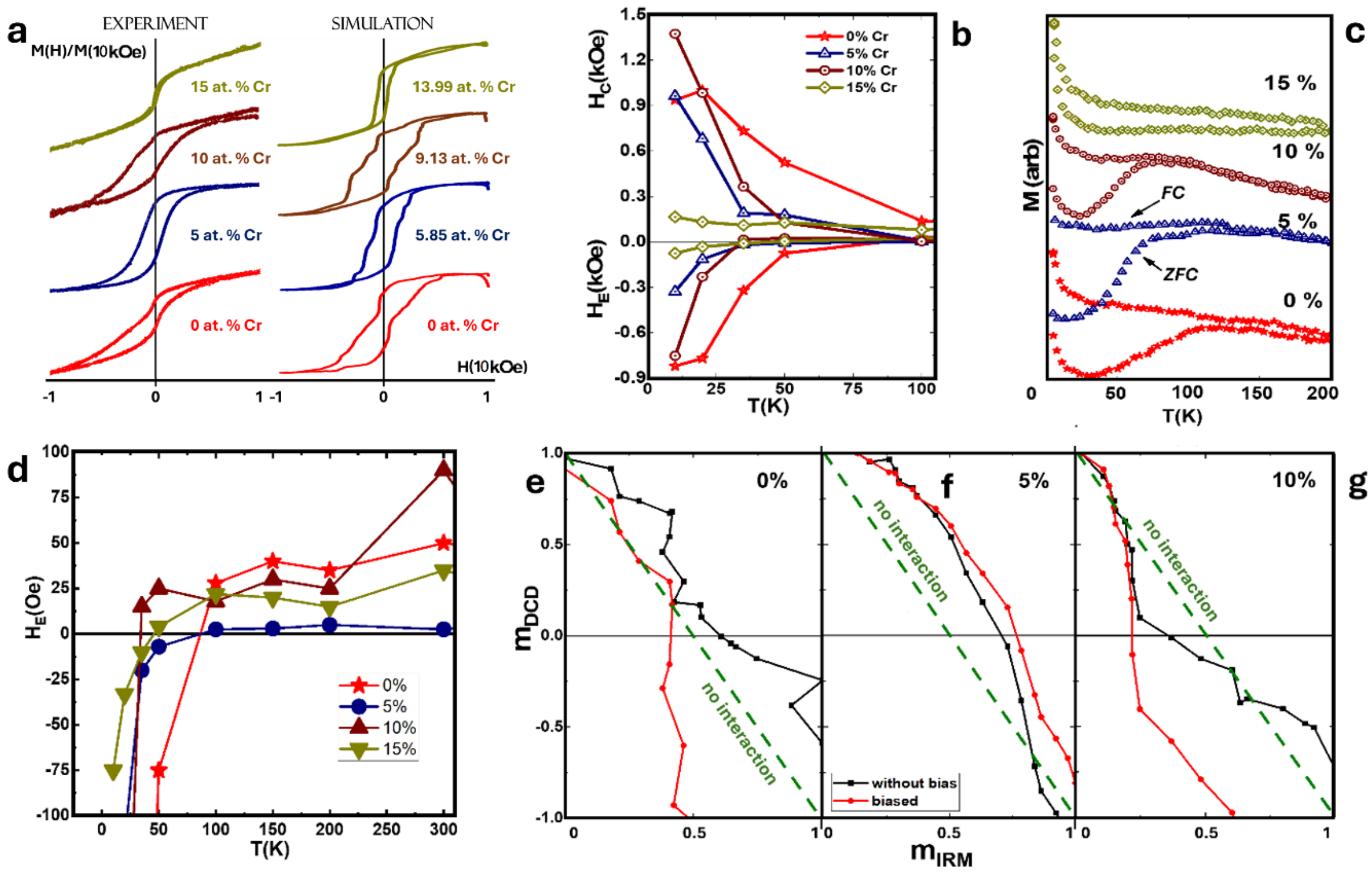

**Figure 3 | Magnetic exchange bias in C/S-NPs. a)** *M-H* loops measured at 10 K under FC conditions (at fixed field of 10 kOe) for all $Ni_{1-x}Cr_x$ ($0 \geq x \geq 0.15$) C/S-NP samples. **b)** Temperature dependence of $H_E$ and $H_C$ curves. **c)** FC and ZFC magnetisation curves (at fixed field of 1 kOe). **d)** $H_E$ vs. $T$ curves around blocking temperatures. **e-g)** Henkel plots, $m_r(H)$ vs. $m_d(H)$, measured at 10 K under biased and unbiased conditions.

To check the strength of the exchange bias at the FM/AFM interface, we deduced the exchange shift and enhanced coercivity using the expressions

$$H_E = (H_{C-} + H_{C+})/2 \quad (1)$$

$$H_C = -(H_{C-} - H_{C+})/2 \quad (2)$$

where $H_{C-}$ and $H_{C+}$ are the coercivities in the negative and positive field directions, respectively [20]. Temperature dependence of the exchange bias field, $H_{EB}$, curves (Figure 3b) indicates that exchange bias appears below 100 K, with $H_E$ values showing maxima of 850 and 750 Oe at 10 K for Ni/NiO and 10 at. % Cr-doped C/S-NPs, respectively, which are quite large for soft FM materials. Remarkably, 5 at. % Cr-doped C/S-NPs show lower $H_E$ values because of the presence of the weaklier–AFM $Cr_2O_3$, which wanes the effective AFM character brought by high–$T_N$ NiO and the corresponding magnetic proximity effect. The presence of interfacial PM $NiCr_2O_4$ has hardly any consequence other than decreasing the exchange bias effect. Coincidently, $H_C$ values (Figure 3b) also start rising around the same temperatures as $H_E$, below 100 K, which is a clear signature of spin-blocking behaviour of the C/S-NPs. This can be confirmed in ZFC magnetisation data, at fixed field of 1 kOe (Figure 3c), where drops in magnetisation occur at similar temperatures (called blocking temperatures, $T_B$). $T_B$ values decrease with increasing Cr doping and, eventually, the 15 at. % Cr sample displays the $1/T$ dependence of magnetisation typical of PM materials. The values of the ratio $|H_E/H_C|$ in all C/S-NPs are less than unity, indicating that not every FM core is fully covered by an oxide AFM shell [21], which is expected due to the high-density coverage of our C/S-NPs.

Another interesting observation is the switchover of negative to positive $H_E$ values below their blocking temperatures, $T_B$, as shown in Figure 3d, which is a typical feature of spin-glassy systems. Zhang *et al.* [22] observed this effect in Fe-film/CoO-NP systems, due to the indirect super-exchange coupling across the FM/SG interface, where Fe moments were coupled in antiparallel fashion with respect to the SG in the CoO. In our C/S-NPs, the presence of multiple competitive FM and AFM phases could lead to SG behaviour [16]. Further, non-zero $H_C$ values at room temperature, which is expected to show SPM properties above $T_B$, indicate that strong interactions dominate over the SPM state.

To elucidate the nature of inter-cluster interactions within $Ni_{100-x}Cr_x$ / NiO-$Cr_2O_3$ C/S-NPs, we produced Henkel plots [23], illustrated in Figure 3e-g, which involve measuring $m_r(H)$ versus $m_d(H)$ curves at 10 K up to ± 10 kOe, under both biased and unbiased conditions. Non-interacting C/S-NPs generally exhibit linear Henkel plots with $m_r(H) = -m_d(H)$, whereas interacting C/S-NPs show curved plots. Curves below the ($m_r(H) = -m_d(H)$) line are attributed to demagnetising interactions, while those above this line represent the magnetising state [23]. For a system of uniaxial particles interacting via magnetic dipolar interaction, this leads to a negative deviation, which increases with the strength of the interaction [23]. The Henkel plots of Ni-NiO C/S-NPs exhibit magnetising interaction (black squares) under unbiased conditions, but under biased conditions they deviate (red squares) below the no-interaction dash line. In the Ni-NiO C/S-NPs, the exchange interfaces between FM and AFM phases have less than 100% contact area [24]. This increases with 5 at. % Cr doping, causing the Henkel plots to cross the no-interaction line under bias. With further increase to 10 at. % Cr doping, the contact area between FM and AFM phases decreases again, causing the Henkel plots to shift to the left of the no-interaction lines. This effect is partially observed under unbiased conditions as well. With greater Cr doping, more Cr is available inside the core, resulting in decreased intra-cluster interactions compared with the direct inter-cluster magnetic interactions, whether they are FM-FM or AFM-FM.

## 5 | Computer Simulations of Structural and Magnetic Properties

To reproduce the magnetic behaviour of experimental C/S-NP ensembles as accurately as possible, fundamental magnetic properties of all observed phases were derived from density functional theory (DFT) calculations and used as input to spin dynamics models corresponding to the experimental samples. Six cases of bulk Ni, NiO, $Ni_{0.875}Cr_{0.125}$ (both FM and PM), Cr, and $Cr_2O_3$ were analysed in accordance with previous experimental and theoretical studies; the respective structural models are presented in **Figures S3-S5**. DFT-calculated ground state lattice constants for the bulk systems, Ni, NiO, Cr, $Ni_{0.875}Cr_{0.125}$ and $Cr_2O_3$, are presented in **Table S2**. Resulting anisotropy and exchange parameters are summarised in **Table S3**, along with the average local magnetic moments per atom as calculated from the DFT simulations and compared with literature values. The magnetic anisotropy constant, $k$, was derived from the magnetocrystalline anisotropy energy, $E_{MAE}$, with a minus sign for configurations with "easy" and "hard" magnetic axes along the [111] and [100] directions, respectively, as described in the **Methods Section**.

Five main configurations of C/S-NP systems were considered for spin dynamics simulations (see **Figure** 4a-e), containing experimentally derived phases (namely, Ni@NiO, $Ni_{0.875}Cr_{0.125}$@NiO with uniform core, $Ni_{0.875}Cr_{0.125}$@NiO with Cr segregated to the edge of the core, and Ni@NiO/$Cr_2O_3$ with Cr segregated to the surface of the NP and oxidised into either one or two $Cr_2O_3$ satellites), roughly corresponding to experimental NPs (see **Methods Section**). Variations to the Ni@NiO configuration were also considered, containing oblong CS-NPs, as will be discussed in a following section. Single-NP hysteresis loops are presented in **Figures S6-S15**, where the coercive field, $H_C$, and the exchange bias field, $H_E$, are defined from equations (1) and (2). The respective values of $H_C$ and $H_E$ for each case of the five considered configurations are presented in Figure 4k-o and Figure 4p-t (also in **Figure S16** for a sixth case of $Ni_{0.875}Cr_{0.125}$@NiO with uniform PM core). Exemplar loops corresponding to the highest exchange bias values are shown in Figure 4f-j. The maximum values of the absolute exchange bias field, $|H_{E,max}|$ and coercivity, $H_{C,max}$, are presented in **Table S4**.

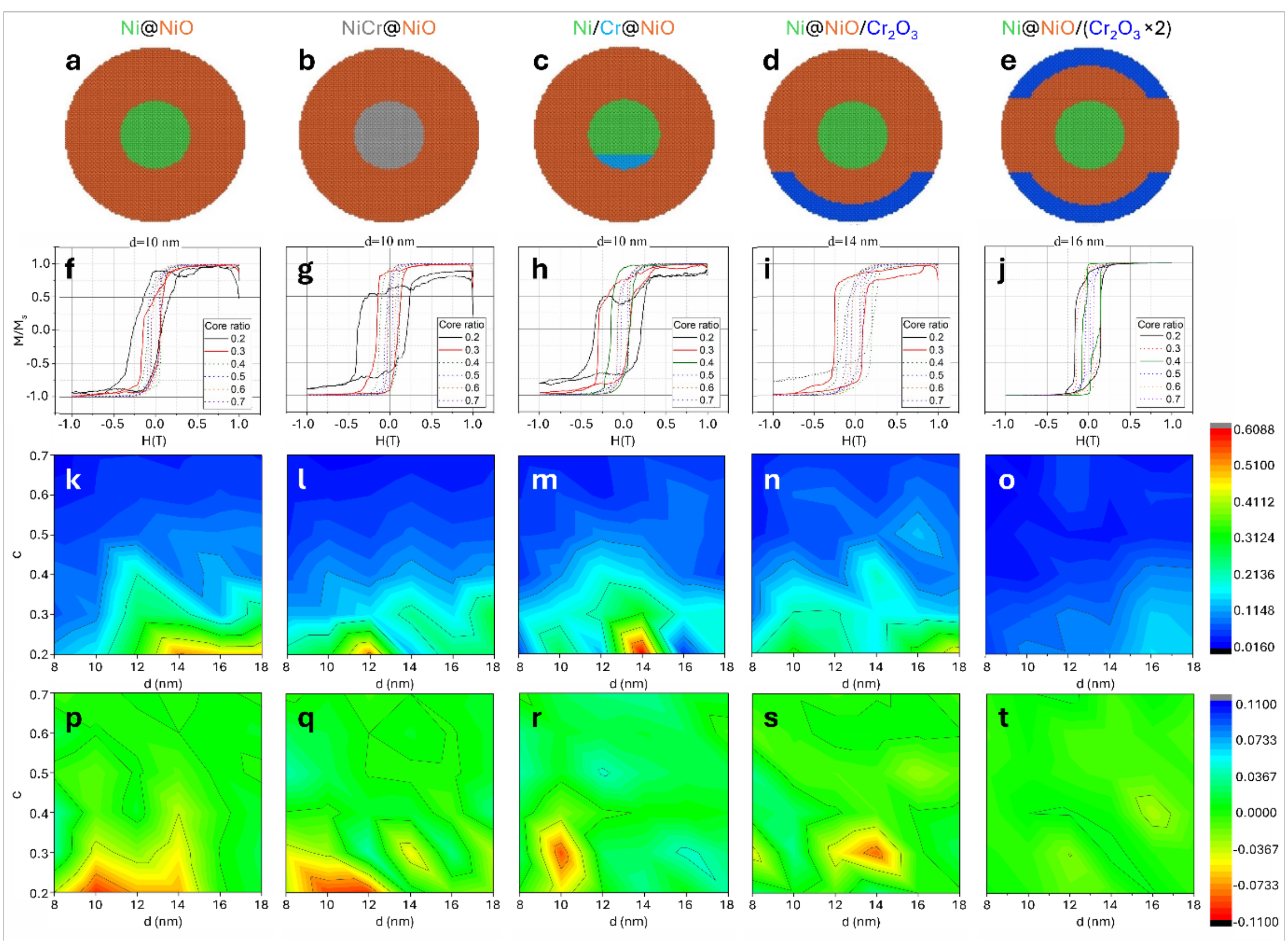

**Figure 4 | Cross-sections of the core-shell NP models used for the spin dynamics simulations and corresponding results. a-b)** Core-shell configuration with two distinct effective materials, used for the cases of Ni@NiO and $Ni_{0.875}Cr_{0.125}$@NiO C/S-NPs. The $Ni_{0.875}Cr_{0.125}$@NiO configuration corresponds to the case where Cr atoms are homogeneously distributed inside the Ni core. Ni and $Ni_{0.875}Cr_{0.125}$ cores and NiO shell are depicted green, grey, and brown, respectively (depicted here is the case where the $Ni_{0.875}Cr_{0.125}$ core is FM). **c)** Core-shell configuration with three distinct effective materials, used for the case of $Ni_{0.875}Cr_{0.125}$@NiO NPs, but with Cr precipitate (cyan) being separated from the Ni core and segregated to a single area near the surface of the Ni core (Ni/Cr@NiO configuration). **d)** Core-shell configuration with three effective materials, used for the case of Ni@NiO/$Cr_2O_3$ NPs, where the Ni atoms are located at the core and NiO comprises the shell, along with a single segregated $Cr_2O_3$ satellite (blue). **e)** Same as d, but with two $Cr_2O_3$ satellites. Images produced with OVITO.[25] **f-j)** Corresponding hysteresis loops from spin dynamics simulations with DFT-calculated parameters for the configurations described in a-e. Full lines represent the cases where $|H_E| > 0.4$ kOe is observed. **k-o)** Contour plots of the coercive field, $H_C$, as a function of the NP diameter, $d$, and core-to-total diameter ratio, $c$, for the five considered NP configurations. **p-t)** Contour plots of exchange bias field, $H_E$, as a function of NP diameter, $d$, and core-to-total diameter ratio, $c$, for the five considered NP configurations.

## 5.1 | Effect of Ni Oxidation (Ni@NiO)

An inescapable observation from Figure 4 is that Ni oxidation leads to higher coercivity and more pronounced exchange bias most of the times. All coercive or exchange bias field maxima were observed in cases where NPs were heavily oxidised, with a core-to-total diameter ratio, $c$, of 0.2 - 0.3. This is in agreement with a previously reported dependence of $H_E$ on the thickness of the FM material, $d_{FM}$, which follows $H_E \propto 1/d_{FM}$ [26]. Seto *et al.*, for instance, did not observe exchange bias experimentally for Ni@NiO NPs with a core diameter greater than the half of that of the whole NP [27]. Also, the increase of the exchange bias was, in most cases, closely followed by an increase of the coercive field, in agreement with previous studies [26,28-30]. The small differences in NP diameter where $|H_{E,max}|$ and $H_{C,max}$ were observed can be attributed to the fact that $H_E$ follows $H_E \propto 1/d_{AFM}$, but $H_C$ is theoretically found to follow $H_C \propto 1/(d_{FM})^n$, where $n \sim 1.0 - 1.5$; the value of $n$ being a subject of scientific debate [28]. The enhancement of coercivity can also be attributed, in general, to the relatively small anisotropy of the AFM material, which caused the irreversible "drag" of the AFM spins, due to the rotation of the FM spins [26]. However, if the anisotropy of the shell is low, the resistance to the rotation of the spins at the core is smaller; therefore, the shift of the negative branch of the hysteresis loop, i.e., the exchange bias, is also reduced [28].

In general, the dependence of the exchange bias and coercivity on the thickness of system components, and, in particular, the AFM oxide region, is quite complicated [31-34], with the exchange bias appearing beyond a critical thickness of the AFM region, $d_{AFM}$, satisfying $d_{AFM}k_{AFM} \geq J_{interface}$ [28]. In all cases considered here, $H_E$ yielded its maximum value for NPs with a diameter close to the middle of the investigated range (8-18 nm) and for very small sizes of the core. This can be attributed to the fact that for small values of NP diameter, $d$, the impact of the fraction of atoms at the interface between core and shell was dominant. There is evidence that spins at the interface contributed, at least partially, to the appearance of exchange bias, as the exchange bias can be intuitively attributed to the additional torque that the AFM spins exert to the FM ones at the interface, which the external magnetic field has to overcome [28,35]. However, for larger values of $d$, the number of the atoms in the bulk part of the NP increased more rapidly and thus their impact became more important (**Figure S17**).

### 5.2 | Effect of Cr (NiCr@NiO and Ni/Cr@NiO)

It is evident that the formation of a homogeneous NiCr core (NiCr@NiO configuration) leads to a slight increase in maximum coercivity, $H_{C,max}$, (from ~ 5.2 to 5.3 kOe), compared with the Ni@NiO configuration, while $|H_{E,max}|$ remains close to 1 kOe in both cases. The impact of Cr segregation within the NP core, which had been previously observed both computationally and experimentally [16], was investigated by considering a pure-Cr precipitate at the edge of the Ni core (Ni/Cr@NiO configuration, Figure 4c). Thus, the size of the FM core was effectively reduced, as Cr was treated as a weak antiferromagnet (Table

S3). In this scenario, coercivity reached its maximum value at a diameter of 14 nm, as in the Ni@NiCr configuration ($H_{C,max}$ at 12-14 nm), and was also enhanced compared with the previous configurations, reaching a value of 0.61 kOe. $|H_{E,max}|$ was also enhanced to ~ 1.1 kOe and was observed for a diameter of 10 nm, as in the Ni@NiO configuration, but for a slightly less oxidised configuration ($c$ = 0.3). The impact of Cr concentration inside the core of the Ni/Cr@NiO configuration (Figure 4c), for an exemplar NP 10 nm in diameter and a core-to-total diameter ratio of 0.2 (i.e., the case displaying the higher exchange bias), is demonstrated in **Figure S18**.

### 5.3 | Effect of Cr Oxidation (Ni@NiO/$Cr_2O_3$)

If surface segregation of Cr occurs faster than the oxidation of the Ni matrix, or if it is enhanced by preferential Cr oxidation, Cr may be oxidised on the NP surface, forming $Cr_2O_3$ satellites (Ni@NiO/$Cr_2O_3$ or Ni@NiO/($Cr_2O_3$×2) configuration, Figure 4d-e). In this scenario, $Cr_2O_3$ was relatively strongly AFM (Table S3); therefore, the AFM character of the shell was effectively reinforced locally, inside this specific region. Thus, segregation of Cr to the outer surface of the NP appears detrimental for its magnetic properties, as both $|H_{E,max}|$ and $H_{C,max}$ were reduced compared to the previous cases ($|H_{E,max}|$ = 0.9 kOe and $H_{C,max}$ = 4.6 kOe, respectively). Maximum exchange bias was observed at a diameter of 14 nm and a slightly smaller shell ($c$ = 0.3), while maximum coercivity was observed for larger NPs ($d$ = 18 nm) and $c$ = 0.2. The impact of $Cr_2O_3$ on the exchange bias was also found detrimental in a previous study regarding Fe NPs embedded in $Cr_2O_3$ matrix [31]. Moreover, AFM NiO NPs were experimentally found to yield almost 5 times higher $H_E$ than $Cr_2O_3$ NPs [36,37].

### 5.4 | Weight-Averaged Hysteresis Loops

Experimental samples consist of numerous NPs, not necessarily identical in terms of size, composition, or chemical ordering, but, rather, with a distribution of values for each of these parameters. Our computational methodology enables to reproduce the magnetic behaviour of experimental samples by calculating hysteresis loops corresponding to NP ensembles. To this end, we created ensemble-simulated hysteresis loops by weight-averaging the contribution of each simulated configuration (two simple case studies of weight-averaged hysteresis loops are presented in **Figure S19**). For the simulated ensemble to correspond to the experimental deposition as close as possible, statistical weights were estimated by making some educated assumptions about the samples, based on experimental observations.

For example, knowing that supersaturation of Cr inside the NP core prevents its surface segregation, we used the PM NiCr@NiO configuration (Figure S16) instead of the FM one to simulate the 15 at. % Cr experimental sample. Further, some approximations were necessary in order to match the simulated composition to that of the experimental sample. An analytical breakdown of the rationale behind our selection of weights is detailed in the **Supporting Information** document; spreadsheets containing DFT-calculated values and facilitating such calculations are also available as **Supporting Information**, so that the interested reader may manipulate the weights and explore their effect on the ensemble loops.

Based on such considerations, we employed spin dynamics simulations to produce exemplar magnetisation loops as they would be generated by NP ensembles with the same structure and size distribution as our experimental deposits. Resultant simulated hysteresis loops are shown in Figure 3a alongside the experimental ones for easy comparison, where it is evident that they reproduce the experimental behaviour adequately. Although no perfect coincidence with the experimental loops is achieved, our simulated ensemble loops clearly indicate the negative exchange bias and "hummingbird" asymmetry of pure Ni@NiO NPs, or the negligible coercivity and non-saturation shown in corresponding experimental loops of highly Cr-doped C/S-NPs, indicating that our methodology can indeed yield valuable information about such magnetic systems.

In particular, the DFT-relaxed ground state of a Ni@NiO core-shell NP with a diameter of 3.2 nm is shown in **Figure 5**a. The NP presents a slight elongation along the [111] plane; also the crystallinities of both the core and the shell are preserved. The atomic magnetic moments of Ni atoms at the core appear pronounced in the core-shell NP configuration compared with their respective value for the bulk, as demonstrated in Figure 5b. From Figure 5c, it is also evident that the core is compressed, because the expected coordination value of *fcc* Ni, which is 12, appears at a cutoff radius of 3.2 Å, shorter than the calculated lattice constant of bulk Ni (~ 3.5 Å) with the same parameters. This effect corroborates the elongation along the [111] plane and can be associated with the aforementioned increase of the atomic magnetic moments. Further, the coordination number decreases after 6 Å, indicating that the shell is strained and the strain is increasing towards the surface of the NP. A sudden drop of the coordination number is also observed at 8 Å, which is close to the core-shell interface, followed by a peak at 9 Å. This indicates a reduction of the Ni content close to the interface, which leads to a subsequent increase in the following region. As a result, to obtain the simulated ensemble loop of Figure 3a we included slightly elongated versions of the configuration of Figure 4a (see Figure 5d; corresponding individual NP hysteresis loops are shown in Figures S12-S14).

In closing, it should be stressed that “ensemble behaviour” does not mean “collective behaviour”. In other words, our ensemble-simulated loops capture the average magnetic behaviour of a large number of isolated C/S-NPs as fabricated by gas-phase synthesis, but ignore any potential magnetic interaction between such NPs. The latter effect, which is known to greatly affect the collective magnetic behaviour of nanoparticulated samples, is left to be explored in a following study, because it would require an immense expansion of the configurational space, which would gravely complicate our discussion. Hence, in this study, we outline the framework of an investigation protocol that can be readily refined and vastly improved for future studies.

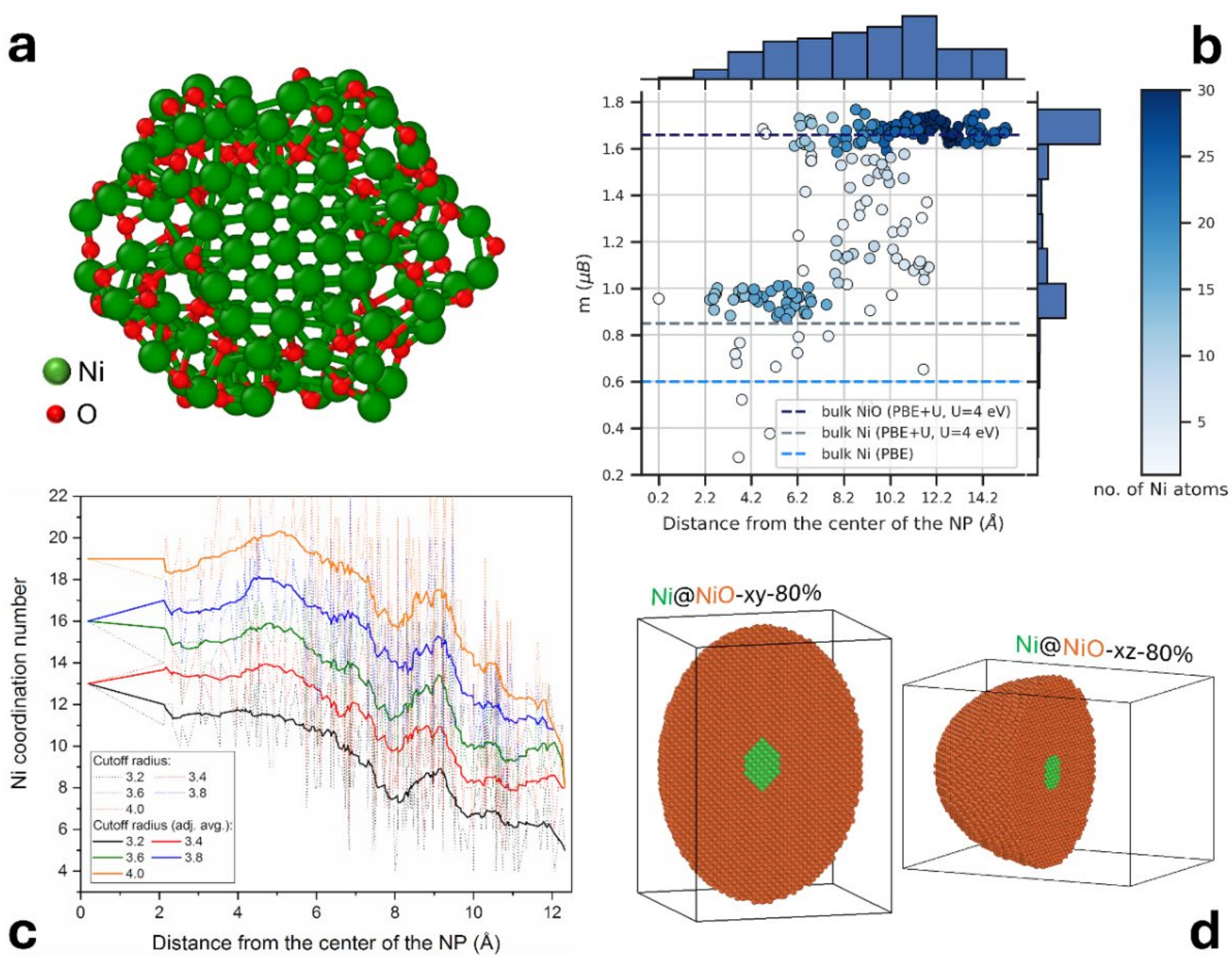

**Figure 5 | Structural and magnetic configuration of a relaxed Ni@NiO C/S-NP. a)** Structural model of the relaxed ground state configuration for a Ni@NiO C/S-NP of a diameter of approximately 3.2 nm and a core-to-total diameter ratio of 0.5. Image produced with VESTA [38]. **b)** Magnitude of the atomic magnetic moments of Ni atoms, *m*, as a function of their distance from the centre of the NP and their respective marginal histograms. Colour coding corresponds to the number of observations on a 2 × 0.2 rectangle in the figure. The atomic magnetic moments of Ni in bulk Ni and NiO are presented with dashed light and dark blue horizontal lines, respectively. **c)** Coordination number of Ni atoms as a function of distance from the centre of the NP, for cutoff radii ranging from 3.2 to 4.0 Å. Continuous lines correspond to adjacent-averaged values. **d)** Core-shell configurations with two distinct effective materials, used for the cases of Ni@NiO. Both CS-NPs are oblong, as their configurations are elongated along the y- or the z-axis. Images produced with OVITO [25].

## 6 | Conclusion

We demonstrate that exchange bias in multiphase metal/oxide nanoparticles is governed not simply by the presence of ferromagnetic and antiferromagnetic constituents, but by their spatial distribution, interfacial topology, and phase competition at the nanoscale. Existing experimental studies have largely focused on specific material systems, whereas theoretical approaches typically model idealised structures that do not capture the structural diversity of real nanoparticle ensembles. As a result, a predictive framework linking nanoscale structure, composition and exchange bias has remained elusive. Here, by combining controlled gas-phase synthesis with first-principles-informed spin dynamics, we disentangle the distinct roles of Cr alloying, segregation, and oxidation in Ni–Cr/NiO C/S-NPs.

Chromium confined within the metallic core strengthens interfacial exchange and enhances both exchange bias and coercivity, whereas its segregation to the surface and subsequent oxidation into $Cr_2O_3$ weakens the effective antiferromagnetic coupling. The observed inversion from negative to positive exchange bias, together with the crossover from exchange to dipolar-dominated interactions, highlights the delicate balance between competing magnetic phases in nanoscale heterostructures.

More broadly, our results establish a predictive design principle: exchange bias in complex C/S-NPs can be maximised by engineering core-to-shell geometry and restricting dopant segregation to preserve strong, coherent interfacial exchange. This framework provides a quantitative pathway for designing exchange-coupled nanostructures that transcend the superparamagnetic limit and advance spintronic, memory and electromagnetic functionalities.

## 7 | Methods

### 7.1 | Fabrication of C/S-NPs

$Ni_{100-x}Cr_x$ (with variable Cr amounts, $x$ = 0, 5, 10, and 15) NPs were synthesised by magnetron-sputtering inert-gas condensation (Nanogen50 Source, Mantis Deposition Ltd., U.K.) at a sputtering power of 40 W from nominal compositional alloy targets. To make C/S-NPs, subsequent to the deposition, as-grown NPs were *ex situ* annealed in air at 200 °C for 1 hour.

### 7.2 | Structural and morphological characterisation

The crystalline structure of C/S-NPs was determined by grazing incidence X-ray diffraction (GIXRD, Bruker D8 Discover XRD2 system with Cu Kα X-ray source) at a grazing angle of 0.25°. X-ray photoelectron spectroscopy (XPS) measurements were performed in a separate ultrahigh-vacuum (UHV) chamber with a base pressure of $2 \times 10^{-9}$ mbar, using a Kratos AXIS Ultra DLD Photoelectron spectrometer with an Al Kα anode. Silicon nitride ($Si_3N_4$) membrane TEM grids were used as substrates for low-coverage samples for transmission electron microscopy (TEM) and scanning-TEM (STEM) analysis using a Cs-corrected environmental transmission electron microscope (ETEM, FEI Titan G2 80–300 kV) operated at 300 kV. The composition of the C/S-NPs was analysed by electron energy-loss spectroscopy (EELS) elemental maps using a post column energy filter.

### 7.3 | Field and temperature dependent magnetic properties

Magnetic properties were measured using a Quantum Design physical property measurement system (PPMS). The diamagnetic contribution from Si substrates was subtracted from the *M–H* and *M–T* data by measuring the magnetic susceptibility of the bare Si substrate of known mass. For the exchange bias study, the magnetisation hysteresis (*M–H*) loops were measured at various temperatures (5–400 K) after field-cooling (FC) at fixed field of 10 kOe; procedures started from 400 K for each measurement. For zero-field-cooled (ZFC) *M-T* data, the sample was initially cooled to 5 K at a zero field, and then magnetisation was measured in the presence of a fixed field of 1 kOe upon heating. Subsequently, the field-cooled (FC) *M-T* data was recorded during cooling in the same field. For Henkel plots, the isothermal remnant magnetisation (IRM) and dc demagnetisation remanence (DCD) curves were measured at 10 K up to ± 10 kOe for both, with (10 kOe) and without bias conditions. The IRM curve was obtained starting from a completely demagnetised state by applying a positive magnetic field; the field was then removed, and the remanence [$m_r(H)$] was measured. The process was repeated at increasing fields up to a maximum of ± 10 kOe. The DCD curve was measured by saturating the sample and then measuring the remanence [$m_d(H)$] after applying a reverse field, $H_{rev}$, up to ± 10 kOe.

### 7.4 | DFT setup

DFT simulations were performed by using VASP 6.2.1 [39] with the projector augmented wave (PAW) version of the PBE and LDA pseudopotentials [40-42] and a plane wave energy cutoff of 520 eV. In all cases, spin polarised, non-collinear calculations were used, with spin-orbit coupling, as well as the aspherical contributions due to the density gradient in the PAW spheres and the Kohn-Sham potential being taken into account. The break condition of the self-consistent loop was set at $10^{-8}$ eV. The magnetocrystalline anisotropy energy, $E_{MAE}$, was calculated as the energy difference between the "easy" and "hard" magnetic axis configurations for each system. For the calculation of the magnetic exchange interaction energy, $J$, a simple Heisenberg model was used, according to which, the total energy of a magnetic system per unit cell can be written as:

$$\mathcal{H} = -\frac{1}{2n}\sum_{i,j} J_{ij}\, \boldsymbol{S}_i \cdot \boldsymbol{S}_j \qquad (3)$$

where $\boldsymbol{S}_i$ is the unit vector of the atomic magnetic spin moment in site $i$, $n$ is the number of unit cells in the crystal, and the summation indices range over all the lattice sites. For FM interactions, $J_{ij} > 0$, while for AFM ones, $J_{ij} < 0$. The magnetic exchange interaction energy between nearest neighbours can be then calculated from the energy difference between FM and AFM configurations:

$$N_i\, J = E_{i,\mathrm{AFM}} - E_{\mathrm{FM}} \qquad (4)$$

where $N_i$ is the number of antiparallel next neighbour spin pairs in the $i$-th AFM configuration.

### 7.5 | DFT Simulated Configurations

The six cases of bulk Ni, NiO, $Ni_{0.875}Cr_{0.125}$ (FM and PM), Cr, and $Cr_2O_3$ were taken into account separately, at a different level of theory for each system, in order to achieve agreement with previous experimental and theoretical studies. In particular, for Ni and NiO, calculations were performed on 2 × 2 × 2 supercells, derived from the respective primitive cells of each structure. The plain PBE functional was used for Ni, while the rotationally invariant PBE+U approach of Dudarev *et al.* [43] was used for NiO, along with a U value of 4 eV, applied for the strongly correlated d electrons of Ni. The Brillouin zones of each system were sampled with automatically generated, $\gamma$-centred 7 × 7 × 7 and 5 × 5 × 5 k-point meshes for Ni and NiO, respectively. The $Ni_{0.875}Cr_{0.125}$ system was represented by substituting a Ni atom with a Cr one, in the 8-atom 2 × 2 × 2 supercell used for the case of bulk Ni, with a $\gamma$-centred 7 × 7 × 7 k-point sampling mesh. The plain PBE-GGA approximation was used, as in previous studies in Ni-Cr systems [44-47]. For Cr, a 2-atom conventional unit cell was used, along with a $\gamma$- centred 16 × 16 × 16 k-points sampling mesh and the LDA approximation, which is argued in the literature to yield magnetic moments and lattice constants closer to the experimental results [48-50]. Finally, for $Cr_2O_3$, a $\gamma$-centred 10 × 10 × 10 k-points sampling mesh and the LDA+U approximation were used, with a value of $U = 3.75$ eV for the 3d orbitals of Cr. The respective structural models are presented in Figures S3-S5.

### 7.6 | Ni@NiO NP

A 381-atom, Ni@NiO core-shell, spherical NP configuration was also investigated via DFT simulations (Figure 5a). The rotationally invariant PBE+U approach of Dudarev *et al.*[43] was used, along with a U value of 4 eV for the d orbitals of Ni. The spin orbit coupling and the aspherical contributions due to the density gradient in the PAW spheres and the Kohn-Sham potential were also taken into account in this case. The radii of the core and shell were taken as equal. The total diameter of the NP was approximately 3.2 Å.

### 7.7 | Spin Dynamics Simulation Setup

For the determination of hysteresis loops, spin dynamics simulations were performed with the VAMPIRE 5.0.1 package [51] and the numerical integration of the atomistic Landau-Lifshitz-Gilbert equation:

$$\frac{\partial \boldsymbol{S}_i}{\partial t} = -\frac{\gamma}{(1+\lambda^2)}[\boldsymbol{S}_i \times \boldsymbol{B}^i_{\mathrm{eff}} + \lambda \boldsymbol{S}_i \times (\boldsymbol{S}_i \times \boldsymbol{B}^i_{\mathrm{eff}})] \quad (5)$$

In eq. (5), $\boldsymbol{B}^i_{\mathrm{eff}}$ is the effective net magnetic field on site corresponding to the magnetic spin moment in site $i$, $\gamma$ is the gyromagnetic ratio and $\lambda$ is the phenomenological Gilbert damping parameter. The effective magnetic field on site $i$, $\boldsymbol{B}^i_{\mathrm{eff}}$ is derived from:

$$\boldsymbol{B}^i_{\mathrm{eff}} = -\frac{1}{\mu_S}\frac{\partial \mathcal{H}}{\boldsymbol{S}_i} \quad (6)$$

where $\mu_S$ is the local spin moment and H is the full magnetic Hamiltonian:

$$\mathcal{H} = -\sum_{i<j} J_{ij}\, \boldsymbol{S}_i \cdot \boldsymbol{S}_j - \frac{k}{2}(S_x^4 + S_y^4 + S_z^4) - \mu_S \sum_i \boldsymbol{B}_{\mathrm{app}} \cdot \boldsymbol{S}_i \quad (7)$$

The three terms in eq. (7) correspond to contributions to the total energy due to exchange interactions, anisotropy and applied field. The second term is the expression for cubic anisotropy, with negative values of the anisotropy constant, $k$, corresponding to materials with “easy” and “hard” axes along the [111] and [100] directions, respectively.

Five configurations of C/S-NP systems were considered for spin dynamics simulations. In all cases, materials were represented as distinct and uniform effective media, characterised by their respective exchange and anisotropy parameters, as calculated from the DFT simulations. The first two cases involved Ni@NiO and $Ni_{0.875}Cr_{0.125}$@NiO core-shell configurations as in Figure 4a,b (and Figure S16 for the case of PM $Ni_{0.875}Cr_{0.125}$ core). The third case involved a $Ni_{0.875}Cr_{0.125}$@NiO core-shell configuration, but with Cr being segregated towards the edge of the Ni core (Ni/Cr@NiO, Figure 4c) [16]. In the fourth case, Cr was considered segregated toward the edge of the NP and oxidised into a single $Cr_2O_3$ satellite, forming a Ni@NiO/$Cr_2O_3$ core-shell configuration (Figure 4d) [16]. The total content of $Cr_2O_3$ in the NP was set at 12.5 %. Finally, a fifth case was examined where two such $Cr_2O_3$ satellites formed (Figure 4e). In all five cases, the diameter of the NP configurations, *d*, was varied from 8 to 18 nm with a 2 nm step and the ratio of the core to the total NP diameter, *c*, was varied from 0.2 to 0.7, thus forming a 6 × 6 configuration space.

### 7.8 | Spin Dynamics Parameters

In all spin dynamics simulations, 150,000 steps were used for thermalisation and another 150,000 for averaging, with a time step of 1 fs. The external field range was set at ±10 kOe with an increment rate of 0.025 kOe. Furthermore, a temperature of 10 K was used, along with critical Gilbert damping, for quasistatic hysteresis loops at the ground state. The values of exchange and anisotropy parameters, *J* and *k*, and atomic magnetic moments used for each effective material are presented in Table S3. The exchange coupling at the interface between the core and shell effective materials, as well as between the Ni core and the segregated Cr section (Fig. 4c), was taken as equal to the exchange parameter of the core, $J_{IF} = J_{core}$, and the interaction was considered FM [2,52,53]. The exchange coupling at the interface between NiO and $Cr_2O_3$ was taken as equal to the exchange parameter of NiO, $J_{IF} = J_{NiO}$, and the interaction was also considered FM [54]. The anisotropy at the surface was taken as random [55].

### Acknowledgments

This work was supported by Anusandhan National Research Foundation, India, ANRF: CRG/2023/002577 and Indo-Austria Project by DST India: DST/ 563 IC/Austria/2024/124 (TPN 120470). Computational work was supported by computational time granted from the Greek Research & Technology Network (GRNET) in the national high-performance computing (HPC) facility, ARIS, under the project NOUS (pr010034), and also by use of the Deigo HPC facility at OIST and the Aristotle University of Thessaloniki (AUTh) HPC Infrastructure and Resources. S.G. acknowledges support by the Hellenic Foundation for Research and Innovation (HFRI) under the HFRI PhD Fellowship grant (Fellowship Number: 962). P.G. was supported by funding from the Regional Institute of Applied Scientific Research (IRICA), Universidad de Castilla-La Mancha, through a senior "Beatriz Galindo" Grant. The authors are grateful to OIST Graduate University for providing nanoparticle deposition facilities and to N. Jian and A. Damianidis for technical support with microscopy and the spreadsheets, respectively.

**Conflicts of Interest**

The authors declare no conflicts of interest

**Data Availability Statement**

The data that support the findings of this study are available from the corresponding author upon reasonable request.

**Supporting Information**

Supporting Information is available online.

Supporting Information Document (containing Figures S1 to S20, Tables S1 to S5)

Excel Spreadsheets (Supporting Information_0% Cr; Supporting Information_5% Cr; Supporting Information_10% Cr; Supporting Information_15% Cr)

## Table of Contents

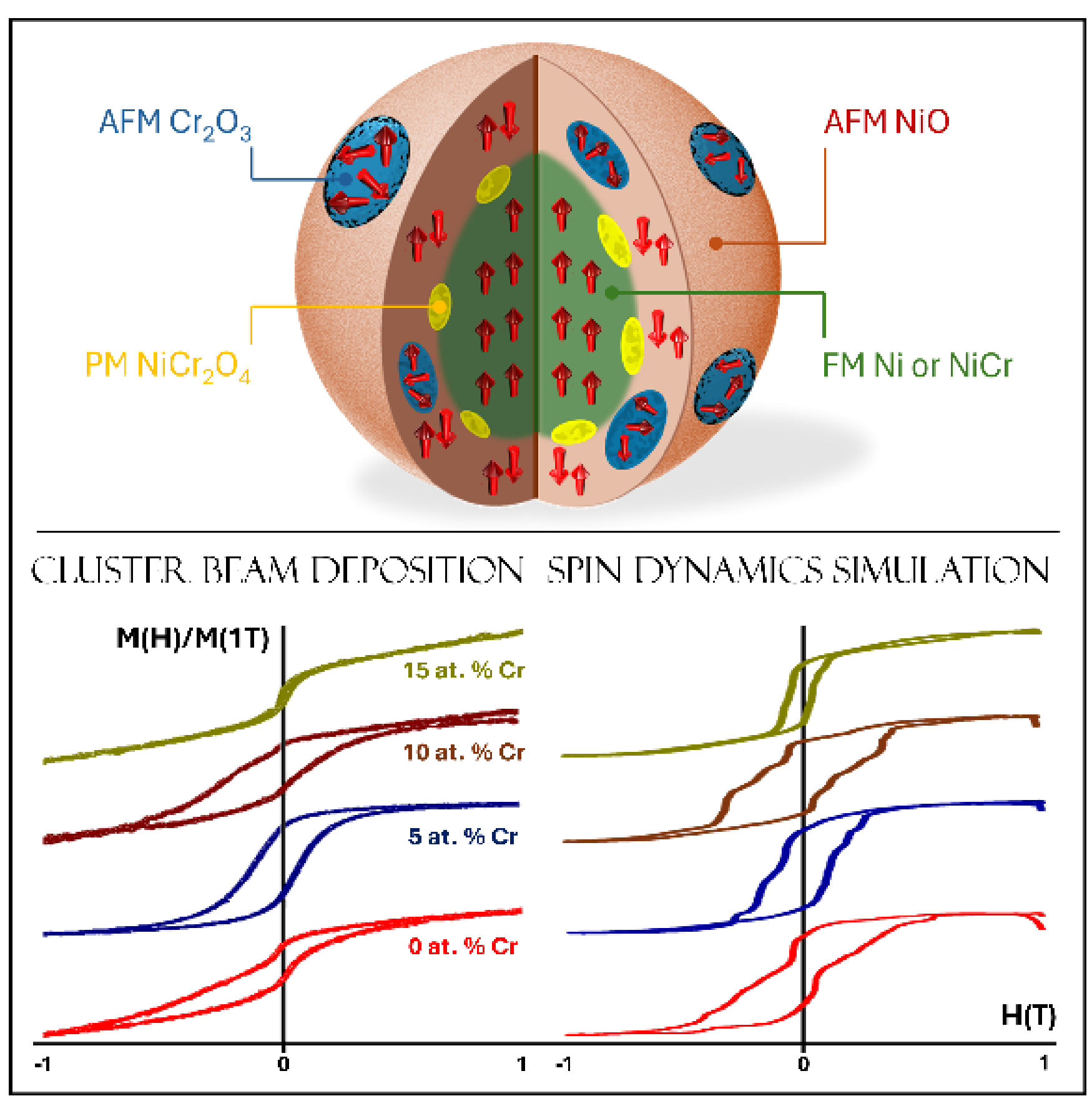

# Supporting Information

## Nanoscale Phase Distribution Governs Exchange Bias in Multiphase Magnetic Nanoparticles

Murtaza Bohra[1], Stefanos Giaremis[2,3], Vidyadhar Singh[4], Joseph Kioseoglou[2,3], Stephan Steinhauer[5] and Panagiotis Grammatikopoulos[6*]

[1]*Ecole Centrale School of Engineering, Mahindra University, Survey Number 62/1A, Bahadurpally Jeedimetla, Hyderabad 500043, Telangana, India*

[2]*School of Physics, Department of Condensed Matter and Materials Physics, Aristotle University of Thessaloniki, 54124 Thessaloniki, Greece*

[3]*Center for Interdisciplinary Research and Innovation, Aristotle University of Thessaloniki, 57001 Thessaloniki, Greece*

[4]*Department of Physics, Jai Prakash University, Chapra, 841301 Bihar, India*

[5]*Department of Applied Physics, KTH Royal Institute of Technology, Albanova University Centre, Roslagstullsbacken 21, 106 91 Stockholm, Sweden*

[6]*Instituto Regional de Investigación Científica Aplicada (IRICA), Departamento de Física, Universidad de Castilla-La Mancha, 13071 Ciudad Real, Spain*

E-mail: p.grammatikopoulos@uclm.es

## 3 | Structure and Chemical Composition of C/S-NPs

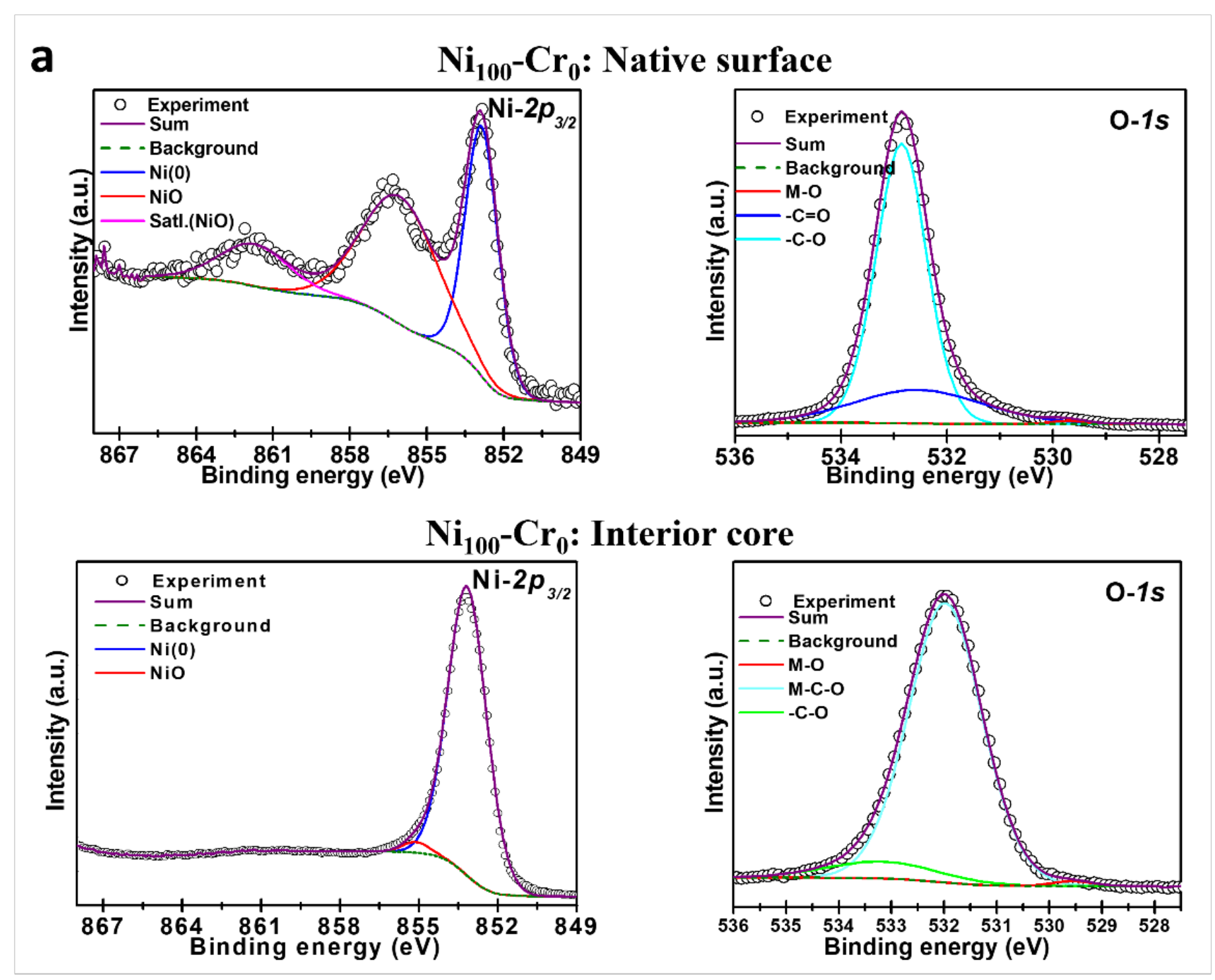

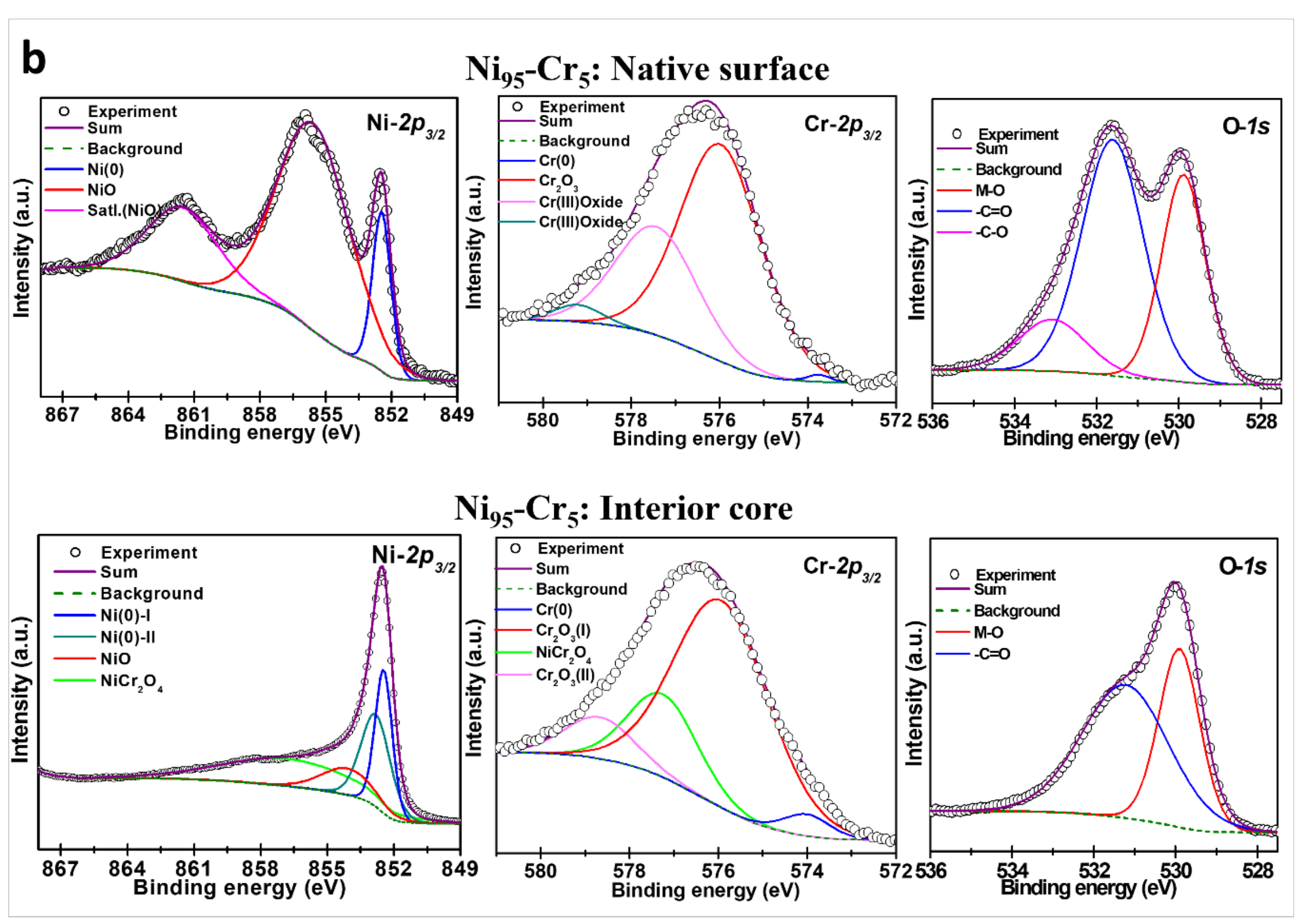

**Figure S1.** Deconvolution of XPS spectra for the determination of the chemical composition of air-annealed C/S-NPs. XPS spectra of **a)** $Ni_{1.0}$-$Cr_{0.0}$ and **b)** $Ni_{0.95}$-$Cr_{0.05}$ CSNs before etching (native surface) and after etching (interior core).

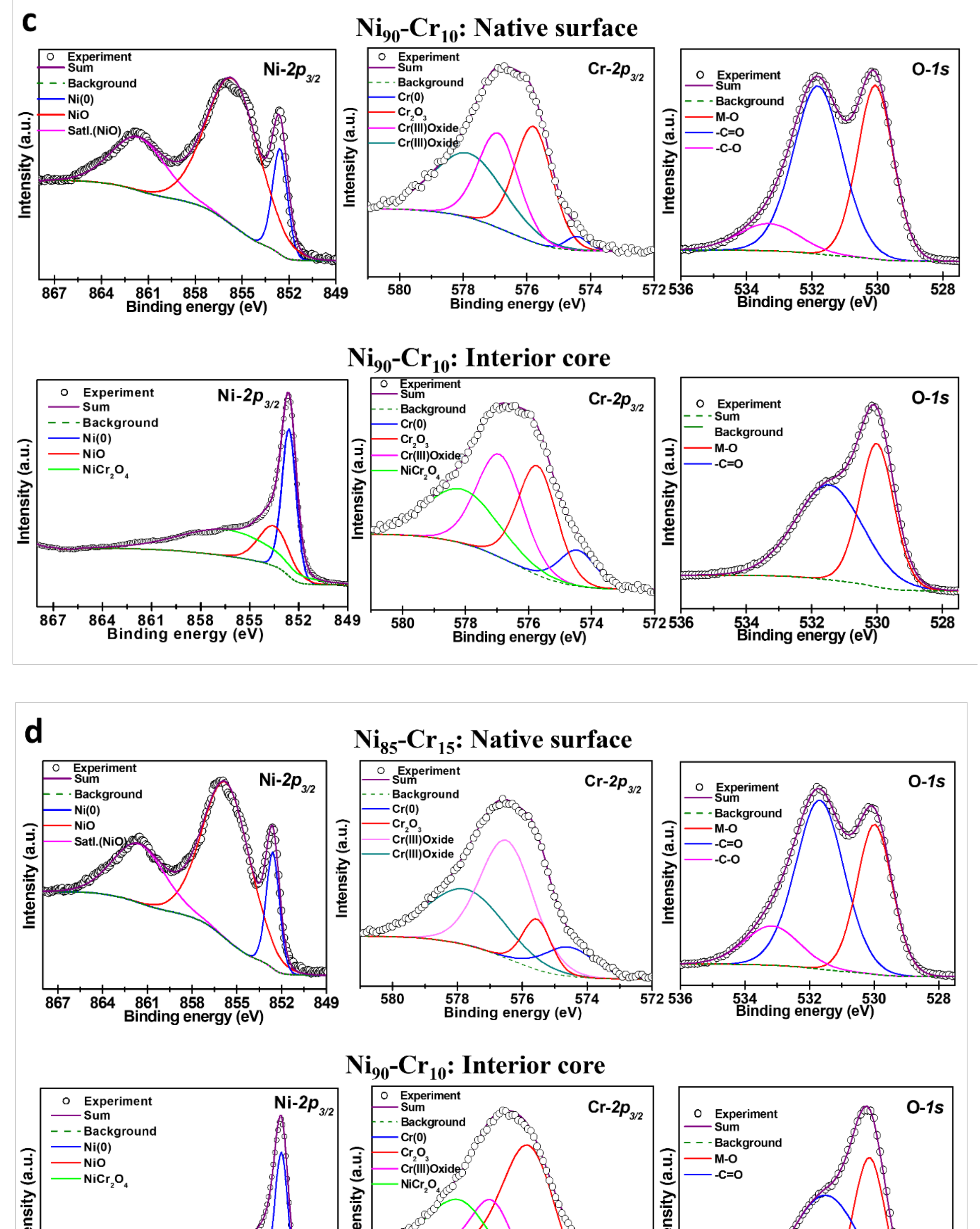

**Figure S1-continued.** Deconvolution of XPS spectra for the determination of the chemical composition of air-annealed C/S-NPs. XPS spectra of **c)** $Ni_{0.90}$-$Cr_{0.10}$ and **d)** $Ni_{0.85}$-$Cr_{0.15}$ CSNs before etching (native surface) and after etching (interior core).

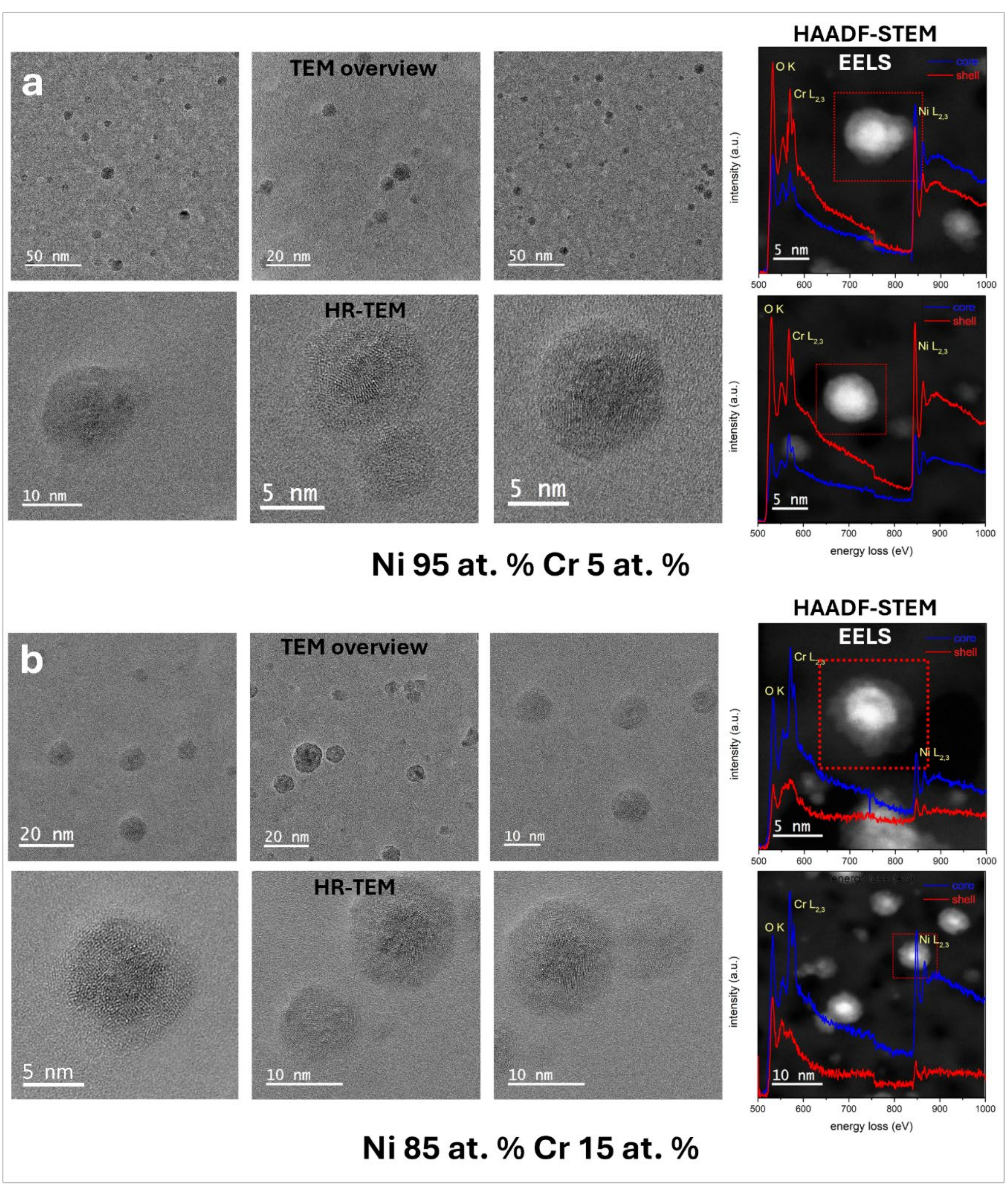

**Figure S2.** Structure and chemical composition of representative $Ni_{1-x}Cr_x$ samples for **a)** x = 0.05 and **b)** x = 0.15. Left panels: TEM survey images and corresponding HR-TEM micrographs of C/S-NPs. Right panels: HAADF-STEM images and corresponding EELS analysis confirming core-shell structures, with cores composed of metallic Ni and NiCr and shells composed of oxides of Ni, Cr and Ni-Cr-O. The relative intensities once more verify the suppression of Cr surface segregation with increasing Cr content.

**Table S1.** Indexing based on lattice spacings from the FFT analysis of the HRTEM images of Figure 2. C/S-NPs have nominal concentrations of **a)** 5 at. % Cr and **b)** 15 at. % Cr.

**a**

| Spot | [1/nm] | d [Å] | possible structure |
|---|---|---|---|
| a | 6.921 | 1.4449 | $Cr_2O_3$ |
| b | 6.5865 | 1.5183 | ~NiO |
| c | 7.173 | 1.3941 | $Cr_2O_3$ |
| d | 8.294 | 1.2057 | NiO |
| r1 | 3.8525 | 2.5957 | $Cr_2O_3$ |
| r2 | 4.757 | 2.2022 | NiO |

**b**

| Spot | [1/nm] | d [Å] | possible structure |
|---|---|---|---|
| a | 4.8865 | 2.0465 | $Cr_2O_3$ / Ni |
| b | 7.92 | 1.2626 | Ni / NiO |
| c | 2.051 | 4.8757 | superlattice |
| r1 | 4.886 | 2.0467 | Ni / NiO |
| r2 | 6.6985 | 1.4929 | NiO |

## 5 | Computer Simulations of Structural and Magnetic Properties

### Effective Material Properties: DFT Calculations

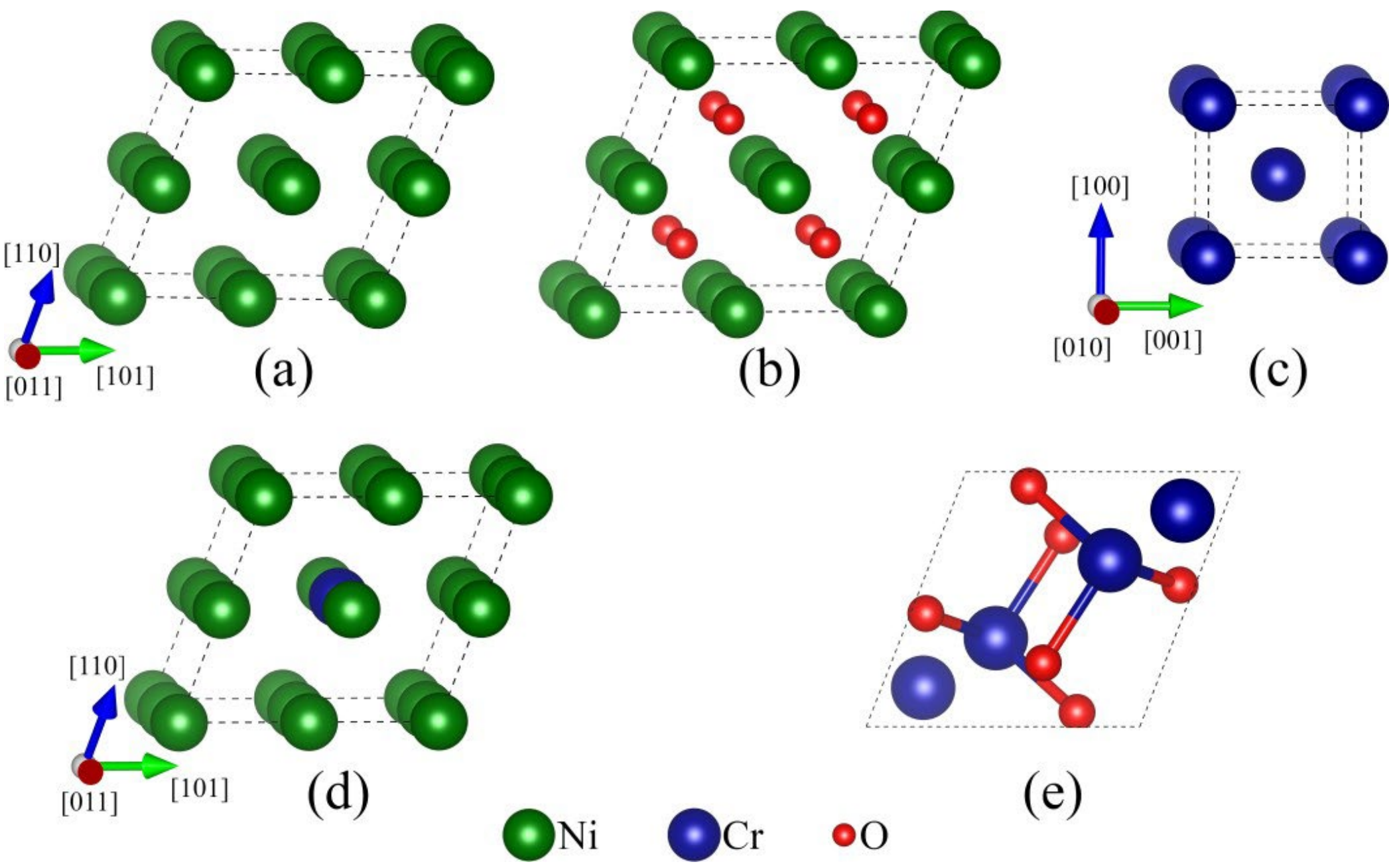

**Figure S3.** Structural models used for the DFT calculations. **a)** 2×2×2 Ni supercell, derived from the respective primitive cell; **b)** 2 × 2 × 2 NiO supercell, derived from the respective primitive cell; **c)** Cr conventional cell; **d)** $Ni_{0.875}Cr_{0.125}$ unit cell, derived from the 2 × 2 × 2 Ni supercell with the substitution of a Ni atom with a Cr one; **e)** $Cr_2O_3$ primitive cell. Images produced with VESTA [S1].

For the $Ni_{0.875}Cr_{0.125}$ stoichiometry case, a Ni atom was replaced by Cr in a 2 × 2 × 2 supercell of the Ni unit cell, leading to a random alloy; this configuration was considered the most reasonable for representing a homogeneous NiCr alloy NP core. The energetically favourable magnetic configuration for this structure was the FM, with the magnetic moment of the Cr atom aligned in parallel with those of the surrounding Ni atoms along the $[11\bar{2}]$ direction. The magnitude of the magnetic moment was enhanced significantly for the Cr atom; it was also increased for the Ni atoms at the corners of the supercell, while it was halved for the rest of the Ni atoms, compared with the corresponding values for each element in their bulk configurations (Fig. S4). This resulted in an average magnetisation of 0.63 μB per atom, which is only slightly higher than the equivalent value for bulk Ni.

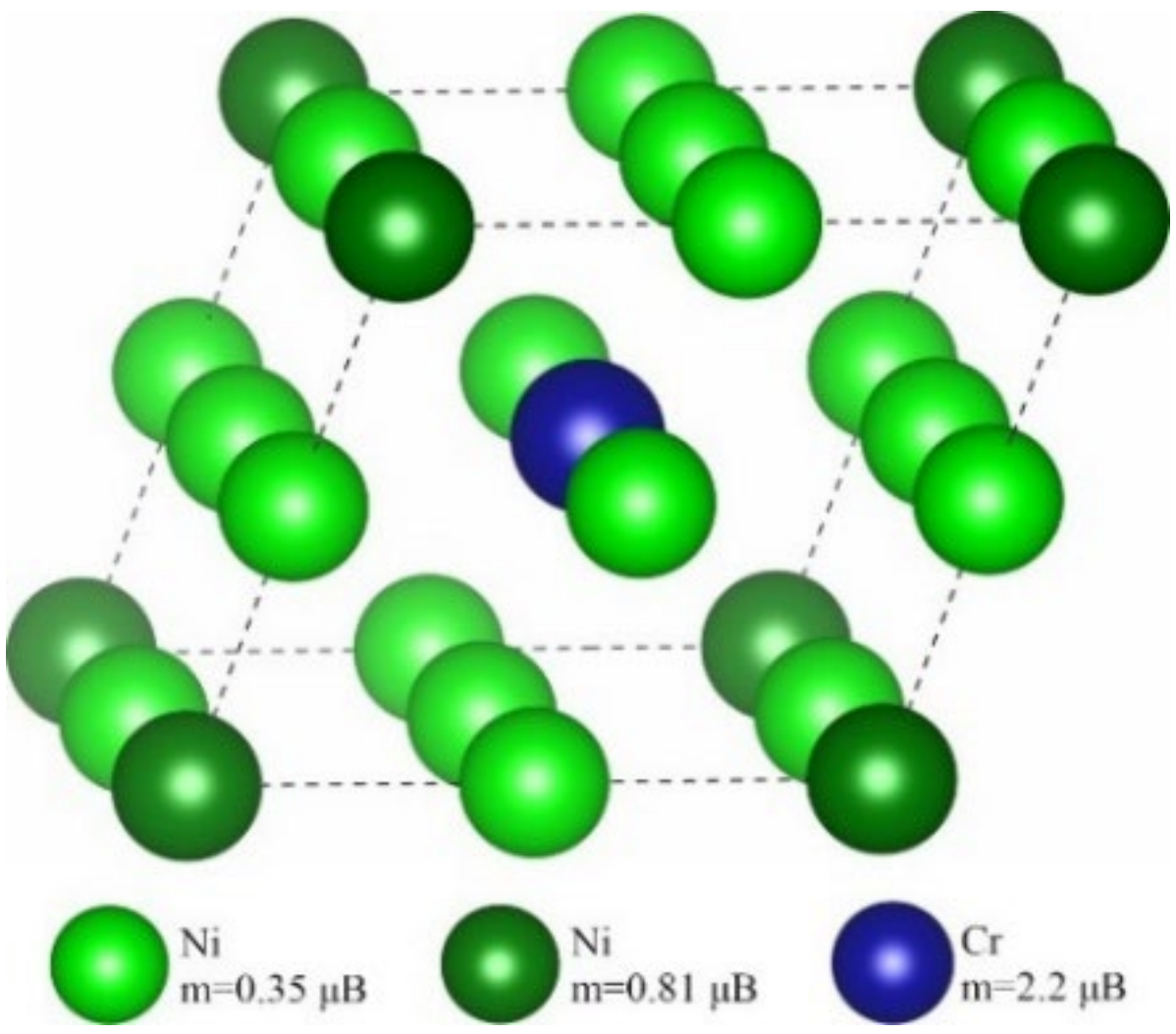

**Figure S4.** Structural model of the unit cell, along with the magnitude of the magnetic moments of each atom, for the $Ni_{0.875}Cr_{0.125}$ system. Image produced with VESTA [S1].

There is evidence in the literature that NiCr with 12 at. % Cr is PM in the bulk [S2]. Furthermore, neutron diffraction studies have demonstrated that short-range ordering occurs in the bulk ground state structure, with intensity maxima observed at $(1\frac{1}{2}0)$ [S3,S4]. By performing DFT simulations with the configuration of Fig. S5, where Cr atoms are replacing Ni ones in the $[1\frac{1}{2}0]$ planes at 12 at. % concentration, the average magnetisation per formula unit drops by one to two orders of magnitude, compared with the case of a random alloy. Therefore, for simulations representing C/S-NPs containing 15 at. % Cr, we replaced the energetically favourable FM $Ni_{0.875}Cr_{0.125}$ with PM $Ni_{0.875}Cr_{0.125}$.

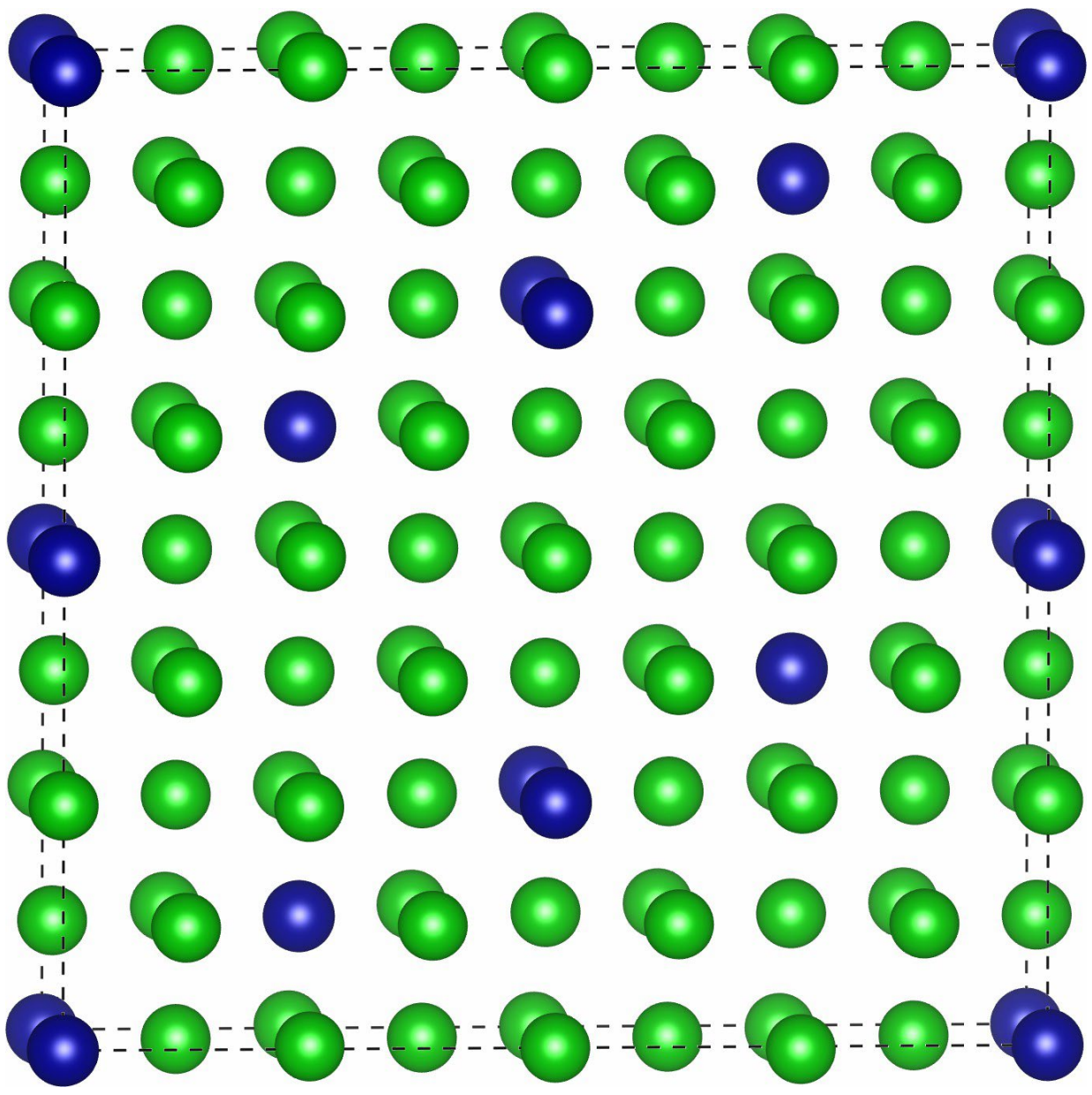

**Figure S5.** Structural model of the PM configuration with the $[1\frac{1}{2}0]$ ordering for the $Ni_{0.875}Cr_{0.125}$ system. Image produced with VESTA [S1].

**Table S2.** DFT-calculated relaxed lattice constants: for bulk Ni, NiO, Cr, $Ni_{0.875}Cr_{0.125}$ and $Cr_2O_3$.

| | Lattice constant (Å) |
|---|---|
| Ni | 3.503 |
| NiO | 4.202 |
| Cr | 2.900 |
| $Ni_{0.875}Cr_{0.125}$ | 3.525 |
| $Cr_2O_3$ | 5.344 |

**Table S3.** DFT output values used a spin dynamics simulations input. Values of exchange, $J$, and anisotropy, $k$, parameters and local magnetic moments, $m$, as calculated from DFT simulations for Ni, NiO, Cr, $Ni_{0.875}Cr_{0.125}$ and $Cr_2O_3$, and subsequently used for the expression of the respective effective materials in the atomistic spin dynamics simulations.

| | $J$ (J/link) | $k$ (J/atom) | $m$ (μB/atom) |
|---|---|---|---|
| Ni | $6.77 \cdot 10^{-22}$<br>$4.49 \cdot 10^{-22}$ [S5] | $-7.82 \cdot 10^{-25}$<br>$-5.61 \cdot 10^{-25}$ [S6] | 0.60<br>0.61 [S7] |
| NiO | $-3.12 \cdot 10^{-21}$<br>$-3.05 \cdot 10^{-21}$ [S8] | $-1.27 \cdot 10^{-24}$<br>$-2.08 \cdot 10^{-24}$ [S9] | 1.66<br>1.79 [S9] |
| Cr | $-9.59 \cdot 10^{-24}$ | $-4.25 \cdot 10^{-25}$ | 0.62<br>0.5 [S10] |
| $Ni_{0.875}Cr_{0.125}$ (AFM)<br>$Ni_{0.875}Cr_{0.125}$ (PM) | $4.02 \cdot 10^{-22}$<br>$5.28 \cdot 20^{-23}$ | $-2.56 \cdot 10^{-25}$<br>$-2.85 \cdot 10^{-25}$ | 0.63<br><0.1 |
| $Cr_2O_3$ | $-2.31 \cdot 10^{-21}$<br>$-2.34 \cdot 10^{-21}$ [S11] | $-8.99 \cdot 10^{-24}$<br>$-8.33 \cdot 10^{-24}$ [S12] | 2.84<br>2.48 [S13]<br>2.76 [S13] |

## Single NP Hysteresis Loops by Spin Dynamics

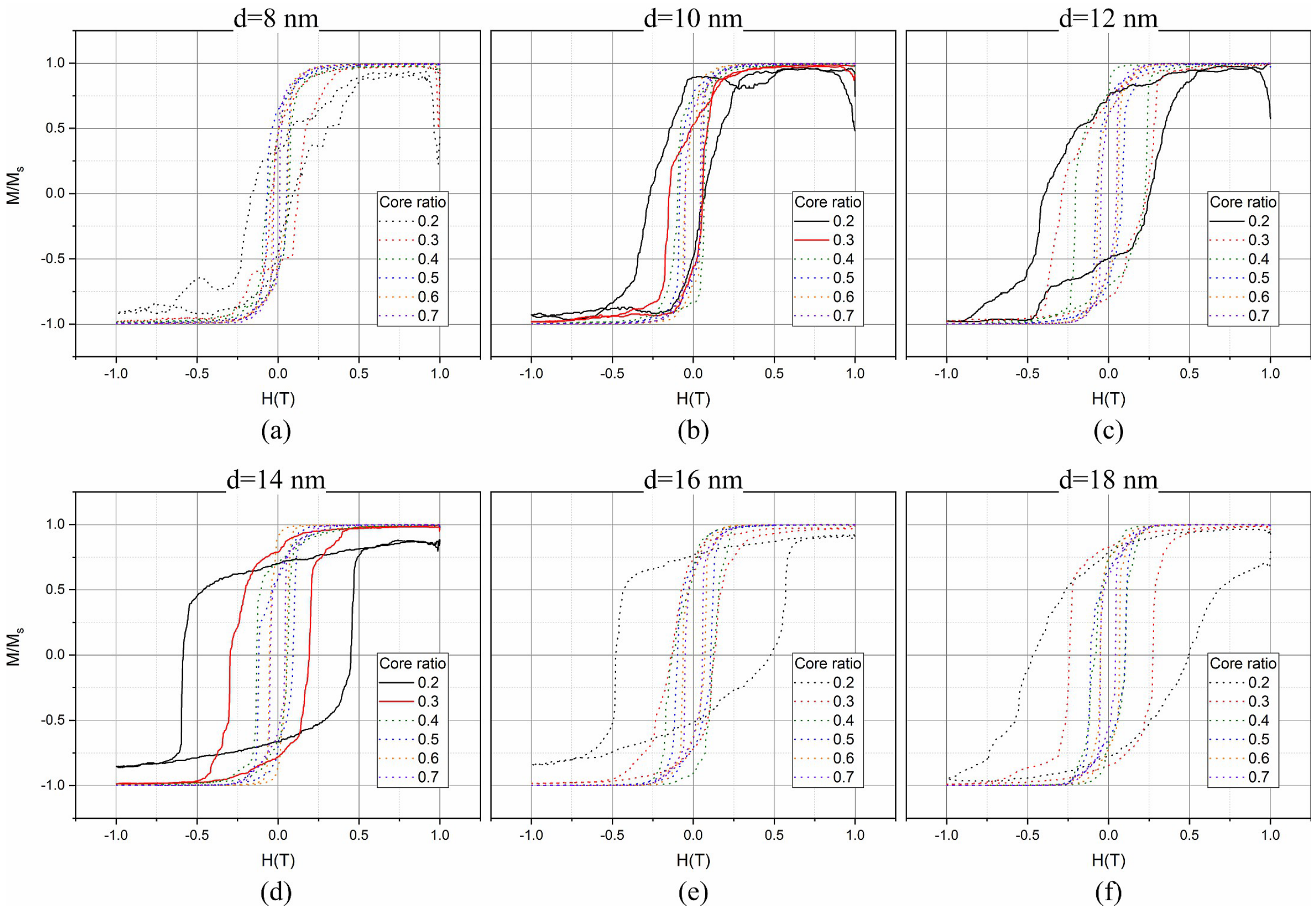

**Figure S6.** Hysteresis loops from spin dynamics simulations with DFT-calculated parameters for the case of Ni@NiO core-shell NPs. Full lines represent the cases where $|H_E| > 0.4$ kOe is observed.

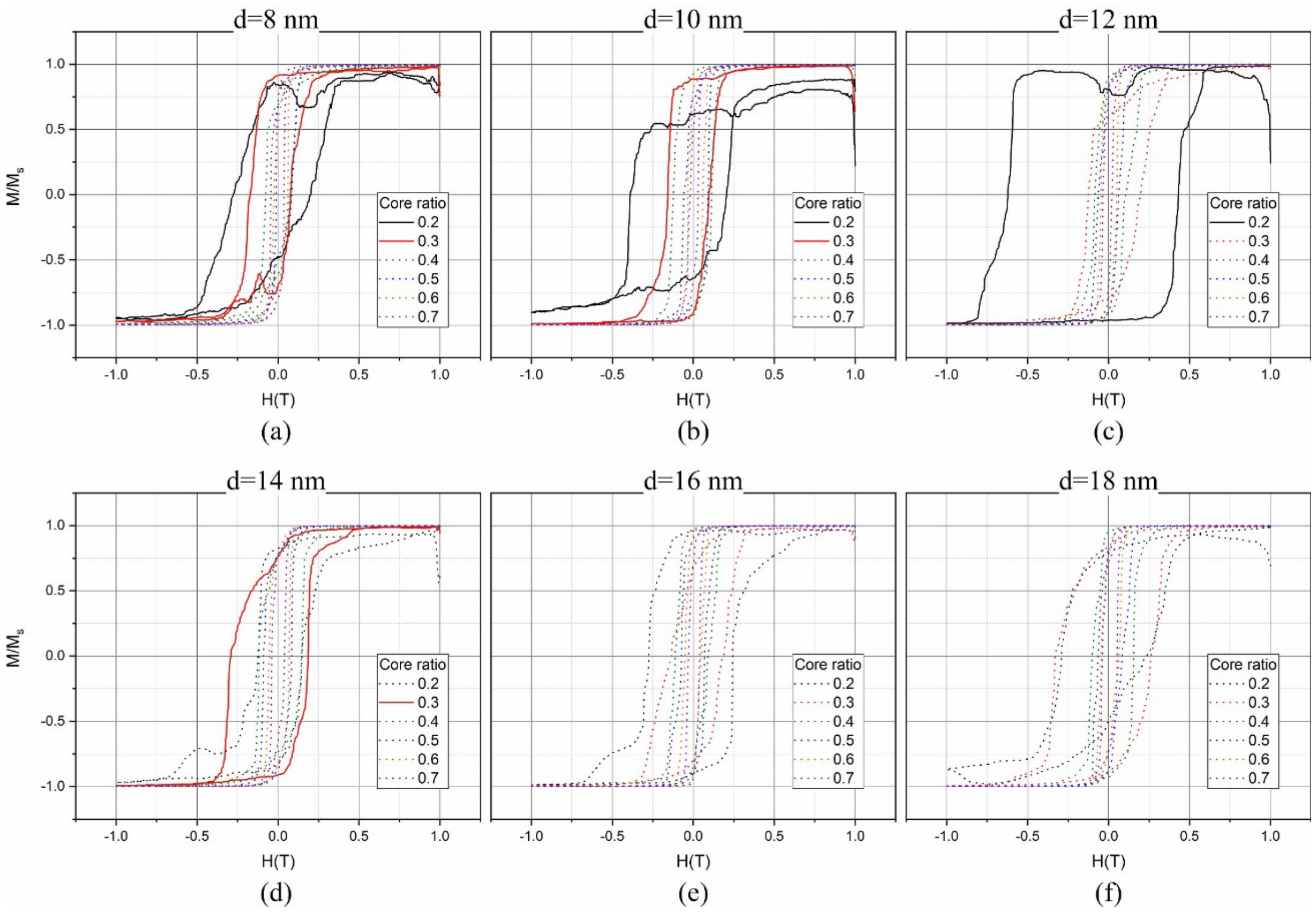

**Figure S7.** Hysteresis loops from spin dynamics simulations with DFT-calculated parameters for the case of $Ni_{0.875}Cr_{0.125}$@NiO core-shell NPs with uniform FM core. Full lines represent the cases where $|H_E| > 0.4$ kOe is observed.

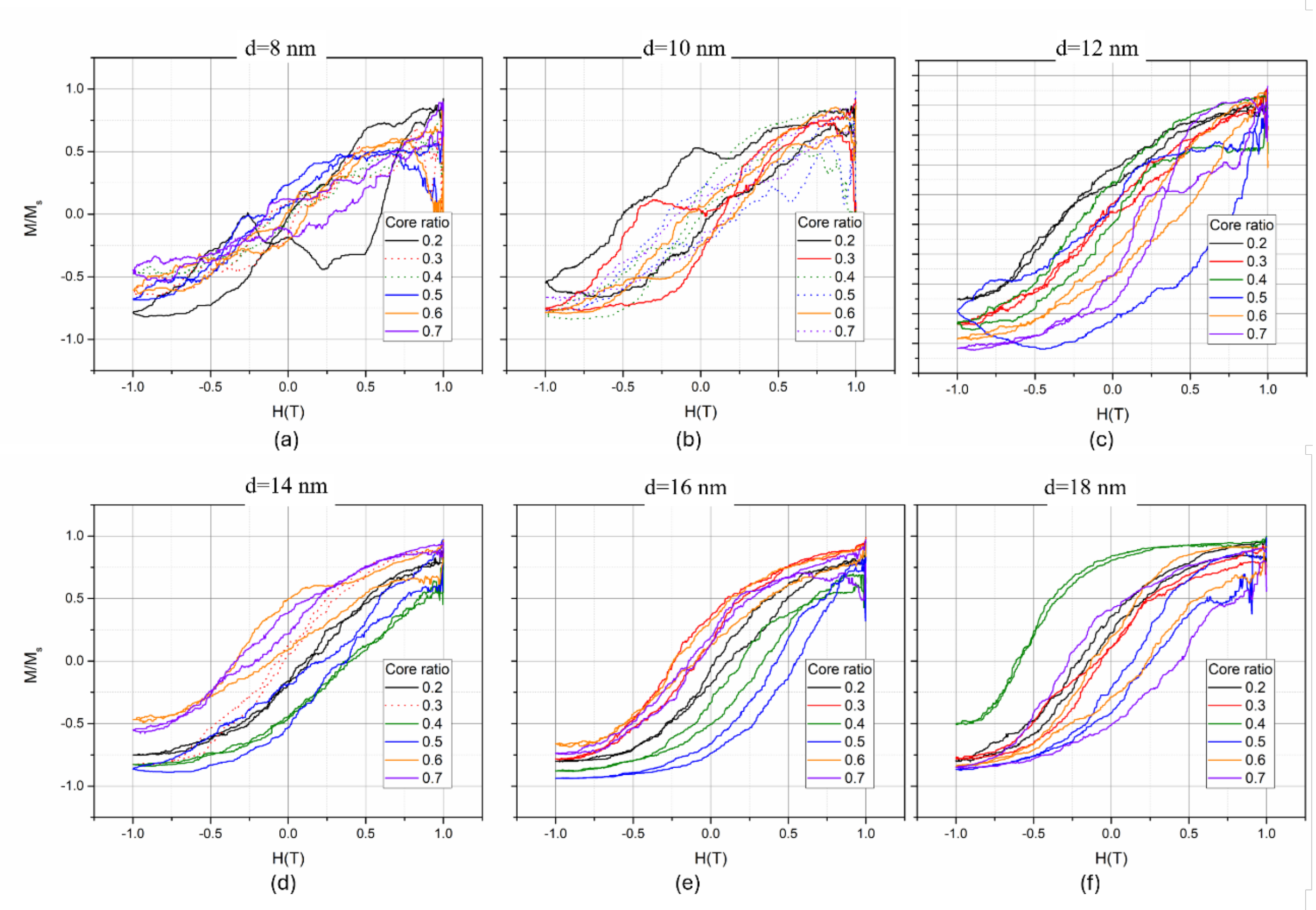

**Figure S8.** Hysteresis loops from spin dynamics simulations with DFT-calculated parameters for the case of $Ni_{0.875}Cr_{0.125}$@NiO core-shell NPs with uniform PM core. Full lines represent the cases where $|H_E| > 0.4$ kOe is observed.

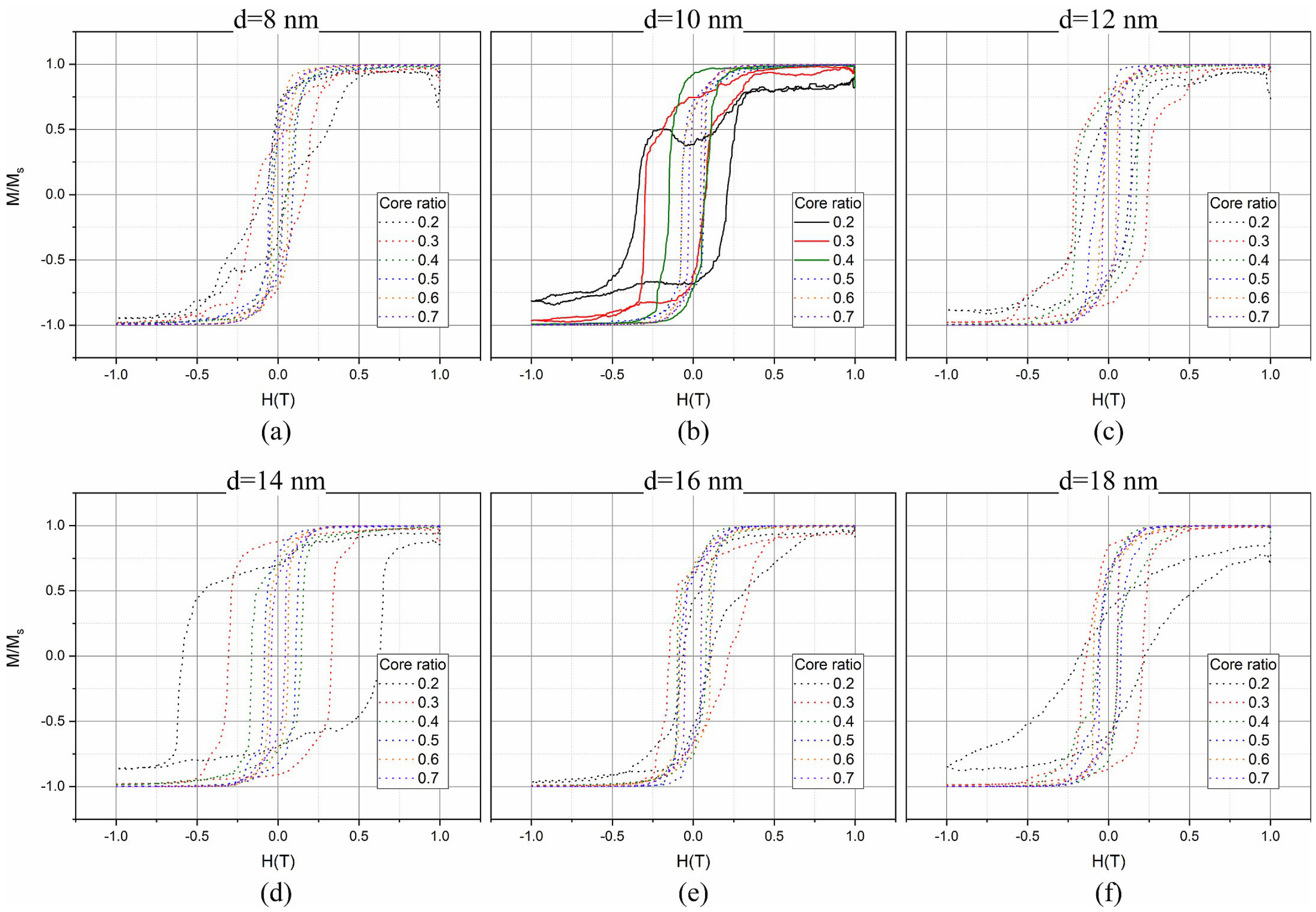

**Figure S9.** Hysteresis loops from spin dynamics simulations with DFT-calculated parameters for the case of $Ni_{0.875}Cr_{0.125}$@NiO core-shell NPs with Cr segregated to the edge of the core (Ni/Cr@NiO). Full lines represent the cases where $|H_E| > 0.4$ kOe is observed.

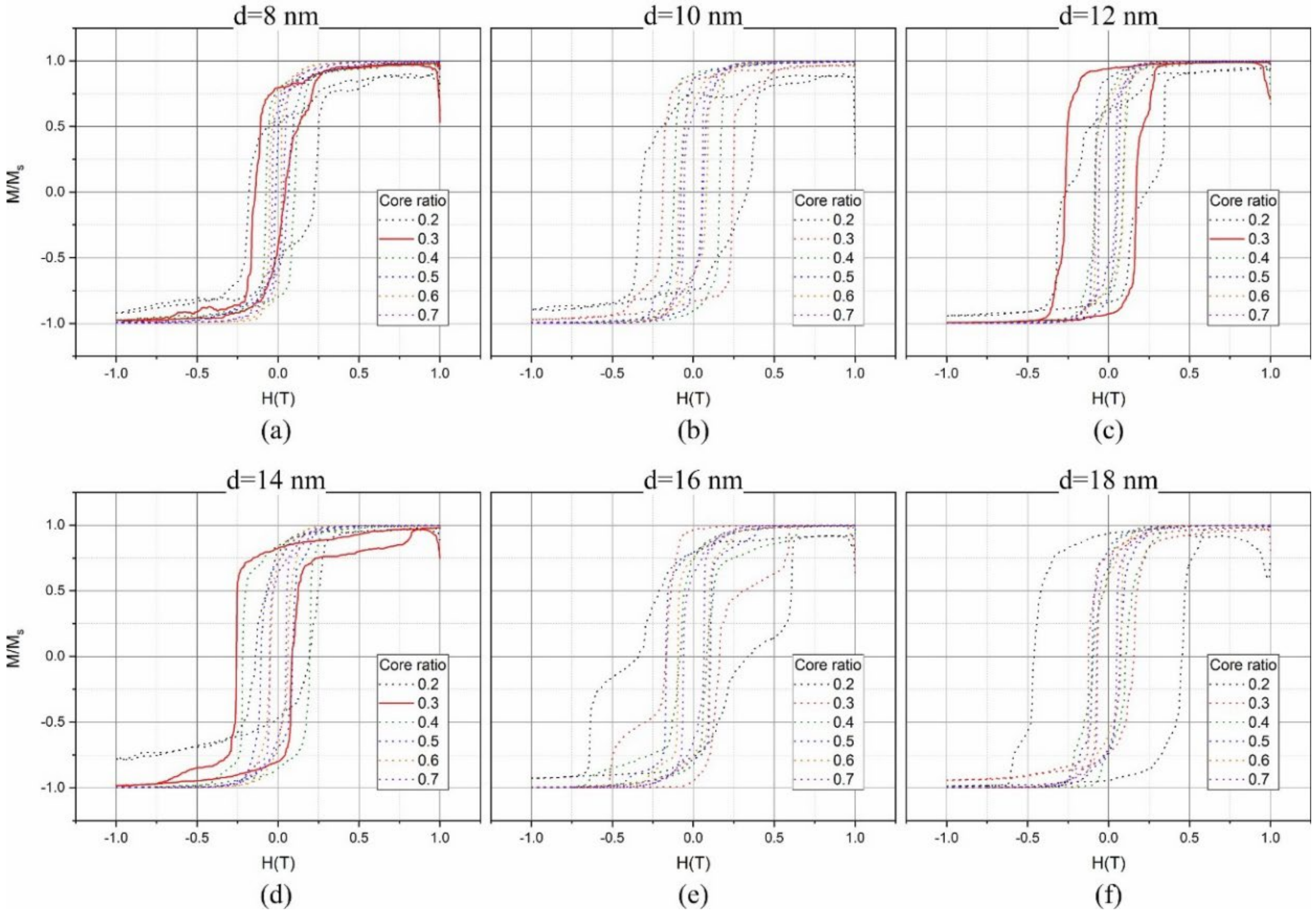

**Figure S10.** Hysteresis loops from spin dynamics simulations with DFT-calculated parameters for the case of Ni@NiO/$Cr_2O_3$ core-shell NPs with Cr segregated to the surface of the NP and oxidised into a single $Cr_2O_3$ satellite. Full lines represent the cases where $|H_E| > 0.4$ kOe is observed.

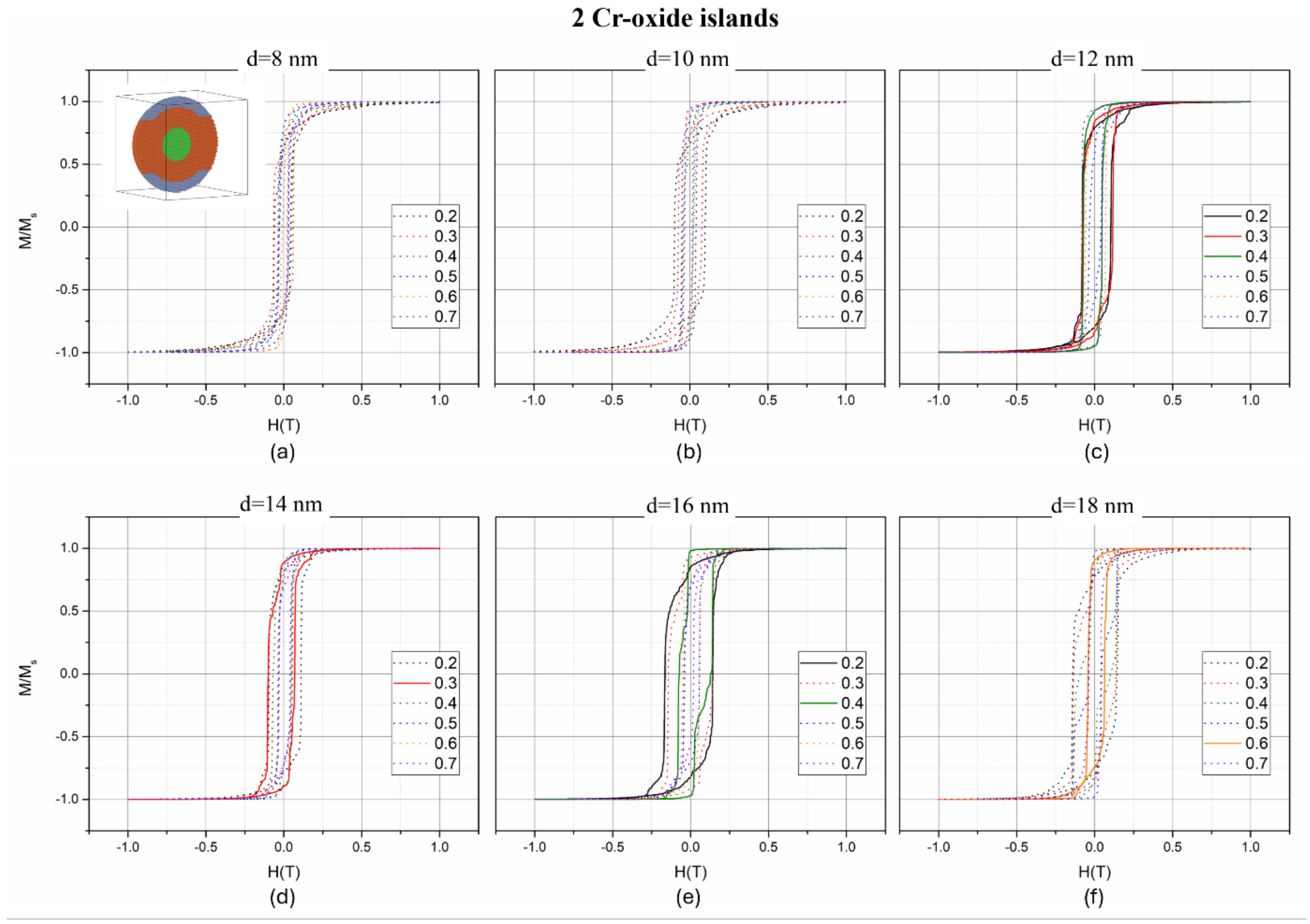

**Figure S11.** Hysteresis loops from spin dynamics simulations with DFT-calculated parameters for the case of Ni@NiO/($Cr_2O_3$ × 2) core-shell NPs with Cr segregated to the surface of the NP and oxidised into two $Cr_2O_3$ satellites. Full lines represent the cases where $|H_E| > 0.4$ kOe is observed.

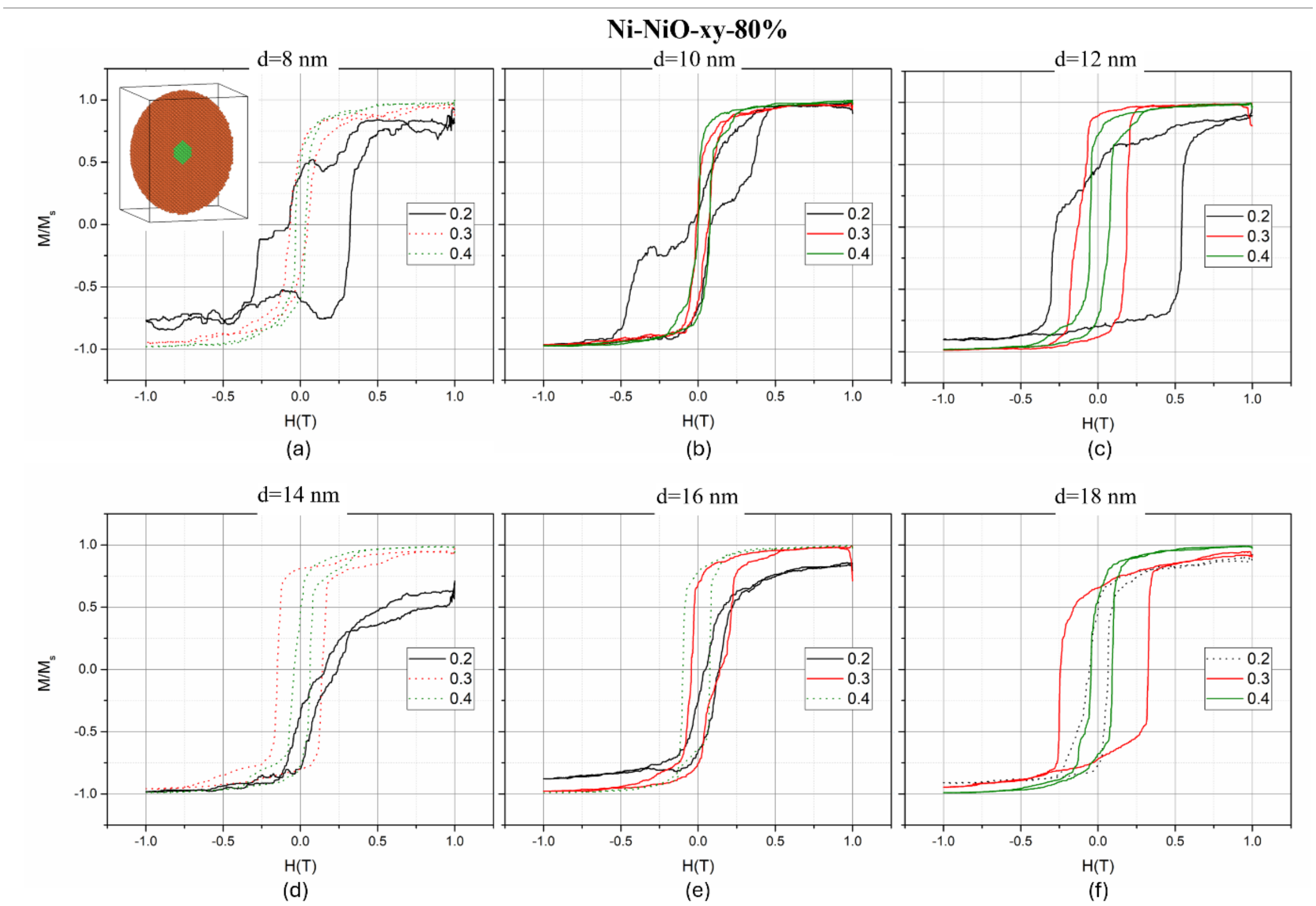

**Figure S12.** Hysteresis loops from spin dynamics simulations with DFT-calculated parameters for the case of Ni@NiO core-shell NPs elongated by 20% along the *z* direction. Full lines represent the cases where $|H_E|$ > 0.4 kOe is observed.

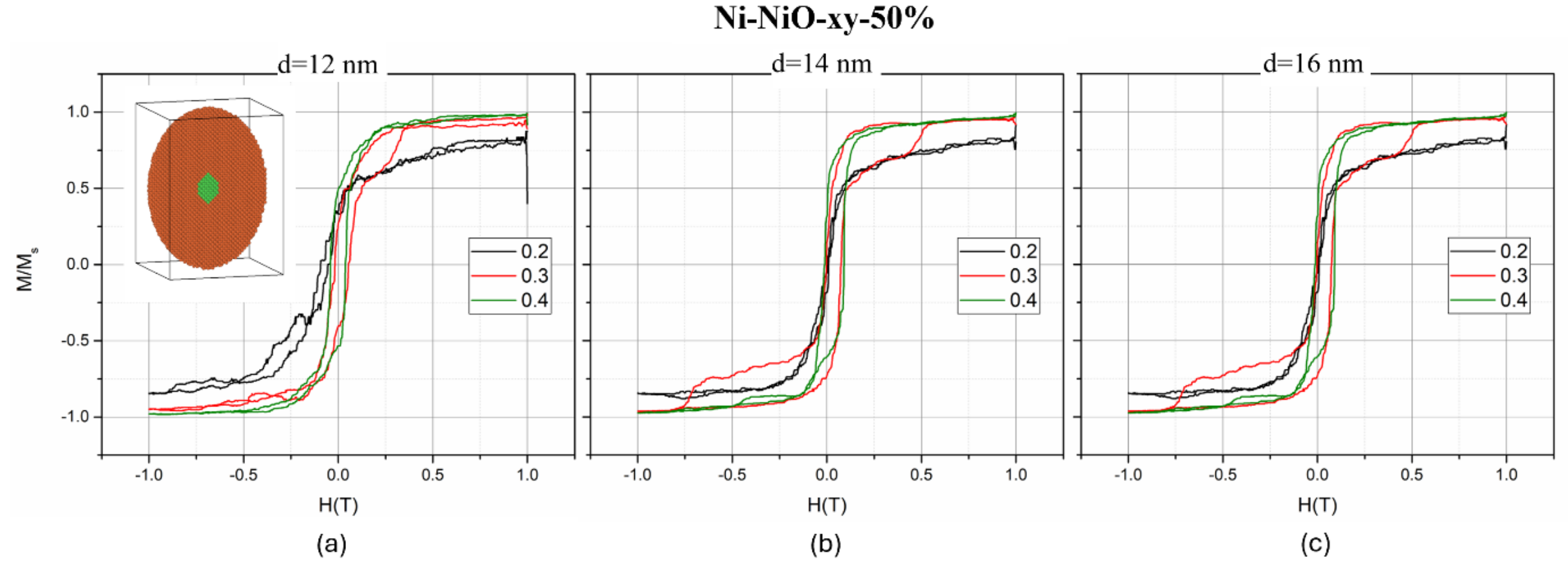

**Figure S13.** Hysteresis loops from spin dynamics simulations with DFT-calculated parameters for the case of Ni@NiO core-shell NPs elongated by 50% along the *z* direction. Full lines represent the cases where $|H_E| > 0.4$ kOe is observed.

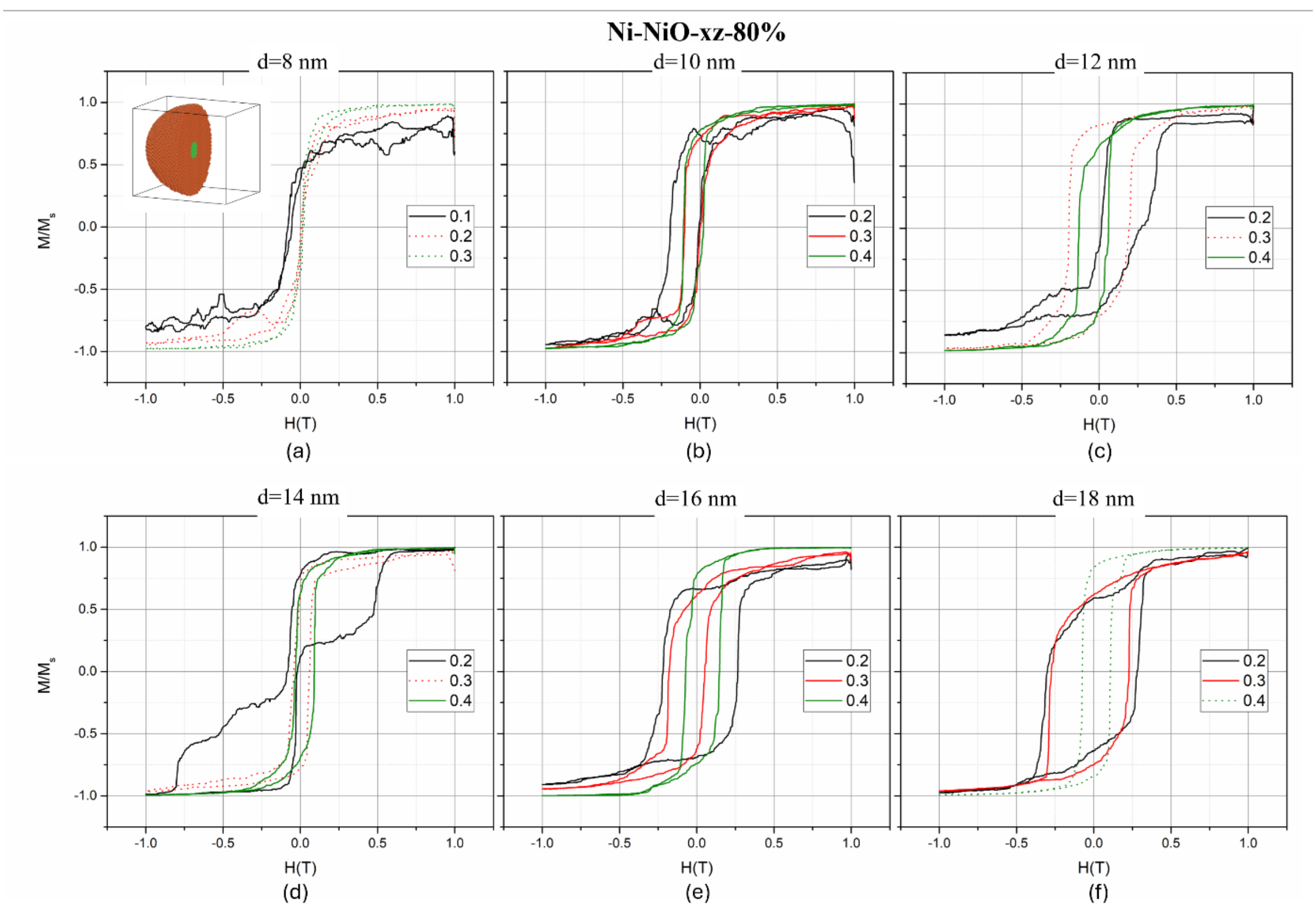

**Figure S14.** Hysteresis loops from spin dynamics simulations with DFT-calculated parameters for the case of Ni@NiO core-shell NPs elongated by 20% along the $y$ direction. Full lines represent the cases where $|H_E| > 0.4$ kOe is observed.

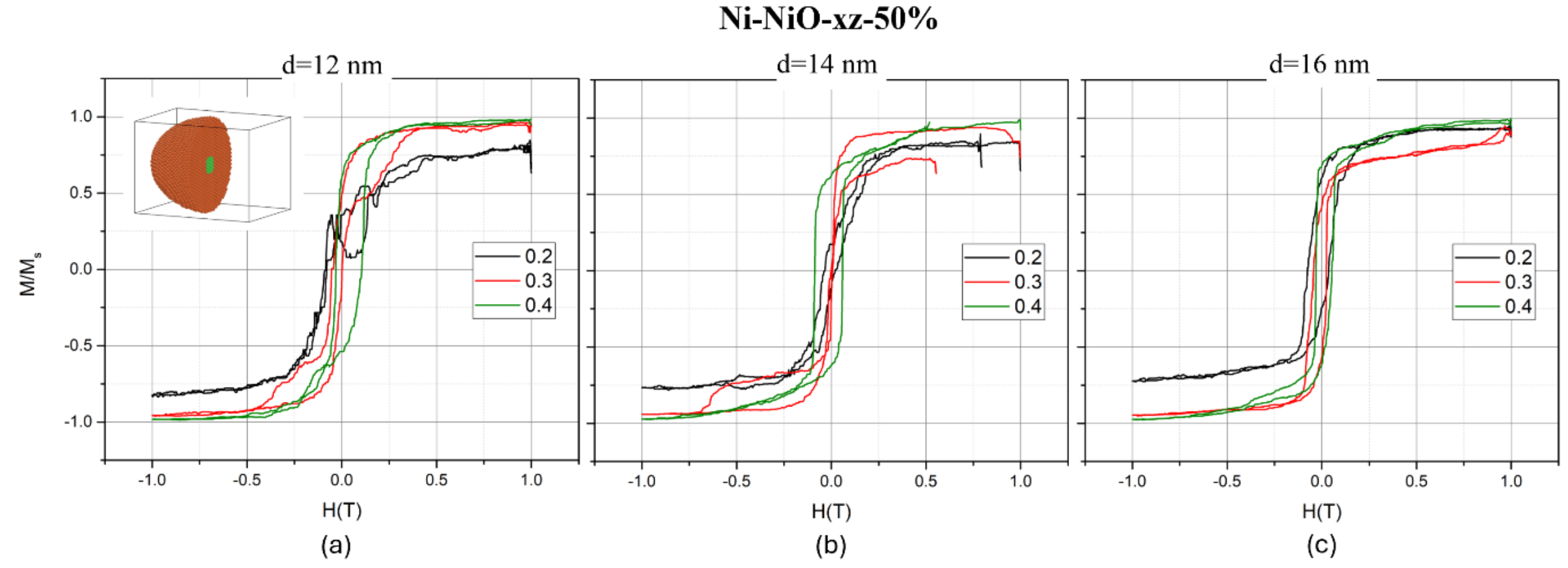

**Figure S15.** Hysteresis loops from spin dynamics simulations with DFT-calculated parameters for the case of Ni@NiO core-shell NPs elongated by 50% along the *y* direction. Full lines represent the cases where $|H_E| > 0.4$ kOe is observed.

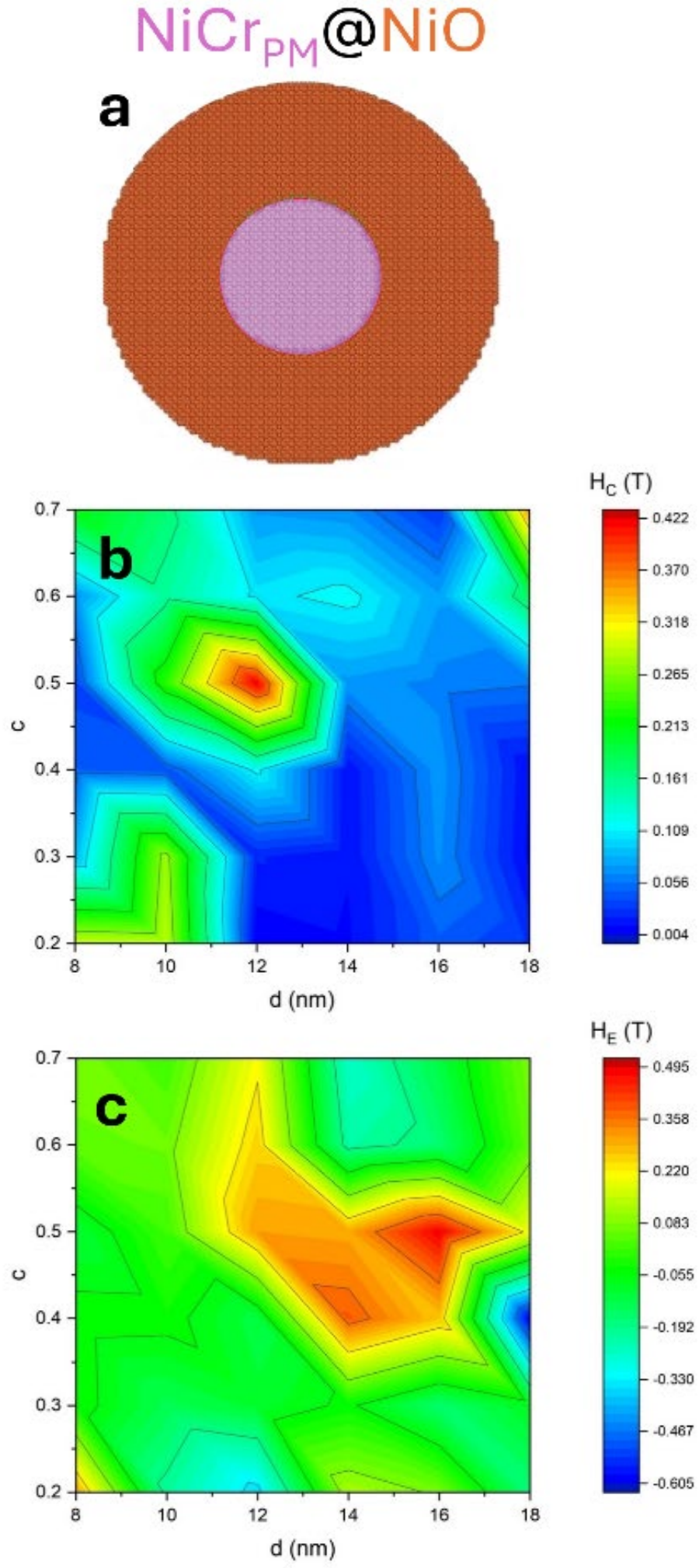

**Figure S16.** Cross-section of the PM-core $Ni_{0.875}Cr_{0.125}$@NiO C/S-NP model used for the spin dynamics simulations and corresponding results. **a)** The configuration corresponds to the case where Cr atoms are homogeneously distributed inside the Ni core, exactly similar to the Figure 4b case; however, here the structure yields PM behaviour. The $Ni_{0.875}Cr_{0.125}$ core and NiO shell are depicted purple and brown, respectively. Image produced with OVITO [S14]. **b-c)** Contour plots of the coercive field, $H_C$, and the exchange bias field, $H_E$, as functions of the NP diameter, $d$, and core-to-total diameter ratio, $c$. Note that due to the PM nature of the core, both fields have distinctly different values of those in Figure 4, hence the decision to plot separately here.

**Table S4.** Calculated values of the maximum absolute exchange bias field, $|H_{E,max}|$, and maximum coercivity, $H_{C,max}$, for each of the four considered NP configurations.

| | $\|H_{E,max}\|$ (kOe) | $H_{C,max}$ (kOe) |
|---:|---|---|
| Ni@NiO | 1.0 | 5.2 |
| NiCr@NiO | 1.0 | 5.3 |
| Ni/Cr@NiO | 1.1 | 6.1 |
| Ni@NiO/$Cr_2O_3$ | 0.9 | 4.6 |
| Ni@NiO/$Cr_2O_3$ × 2 | 0.3 | 1.5 |

**Effect of Ni Oxidation (Ni@NiO)**

The interplay between two mechanisms (interface and volume effects) can be roughly quantified if we consider an effective volume, $V_{\mathrm{eff}}$, defined empirically by:

$$V_{\mathrm{eff}} = \frac{k_{\mathrm{shell}}}{k_{\mathrm{core}}} \cdot S_{\mathrm{interface}} \cdot d_{\mathrm{interface}} - \frac{J_{\mathrm{core}}}{J_{\mathrm{shell}}} \frac{m_{\mathrm{core}}}{m_{\mathrm{shell}}} \cdot (V_{\mathrm{shell}} - V_{\mathrm{core}} - S_{\mathrm{surface}} \cdot d_{\mathrm{surface}}) \quad \text{(S1)}$$

The first term corresponds to the volume of the interfacial area multiplied by the ratio between the anisotropy of the core and the shell and corresponds to the impact of the atoms at the interface and the magnitude of the anisotropy of the AFM material compared with that of the FM. The second term is the excess volume of the AFM material compared to the FM, minus the volume of the surface area, all multiplied by the ratios of the exchange interactions and the magnetic moments between the core and shell. This term can roughly express the impact of the bulk-like atoms on the excess volume of the AFM material and the relative strength of the interactions inside the bulk-like region of the FM material, compared with their shell counterparts. For example, in the case of a simple Ni@NiO core-shell configuration with $c = 0.2$ and $d_{\mathrm{interface}} = d_{\mathrm{surface}} = 1$ nm, the resulting values of $V_{\mathrm{eff}}$ as a function of NP diameter, $d$, are shown in Figure S17.

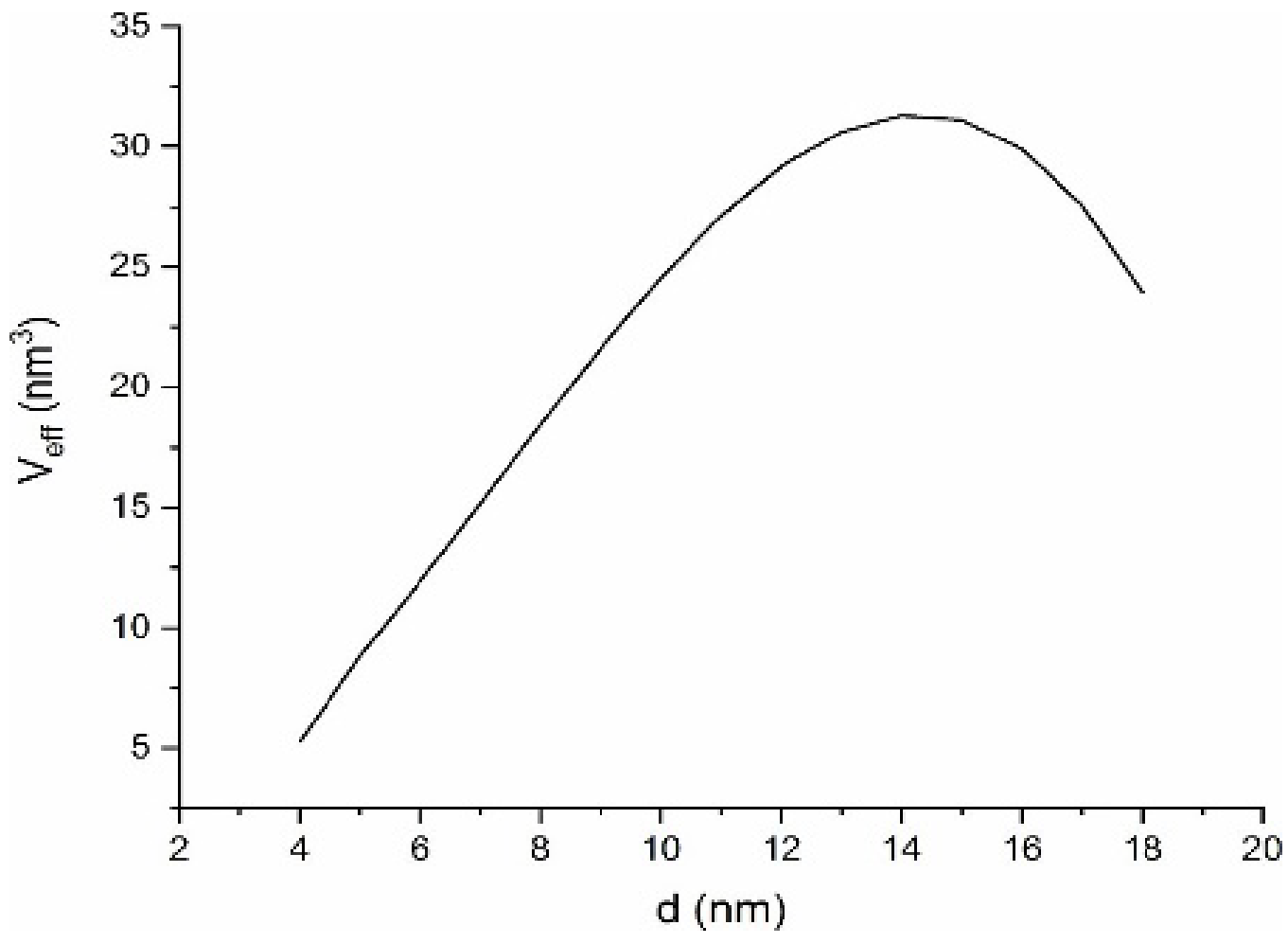

**Figure S17.** Effective volume, $V_{\mathrm{eff}}$, defined by eq. (S1), as a function of NP diameter, $d$, for the case of the simple Ni@NiO core-shell configuration with $c = 0.2$ and $d_{\mathrm{interface}} = d_{\mathrm{surface}} = 1$ nm.

**Effect of Cr (NiCr@NiO and Ni/Cr@NiO)**

Two cases are compared, containing 9.7% and 18.1% Cr at the core, respectively (Figure S18). In the latter case, the enlargement of the Cr precipitate and the corresponding diminution of the remaining Ni core as a result of the higher Cr concentration effectively reduced the volume of the FM domain at the core. Reduced Cr content led to a shift of exchange bias toward a pronounced positive value ($H_E$ = 0.9 kOe, compared with $H_E$ = -0.07 T for the case with increased Cr content), at the cost of reducing coercivity almost to half ($H_C$ = 1.3 kOe, compared with 2.8 kOe).

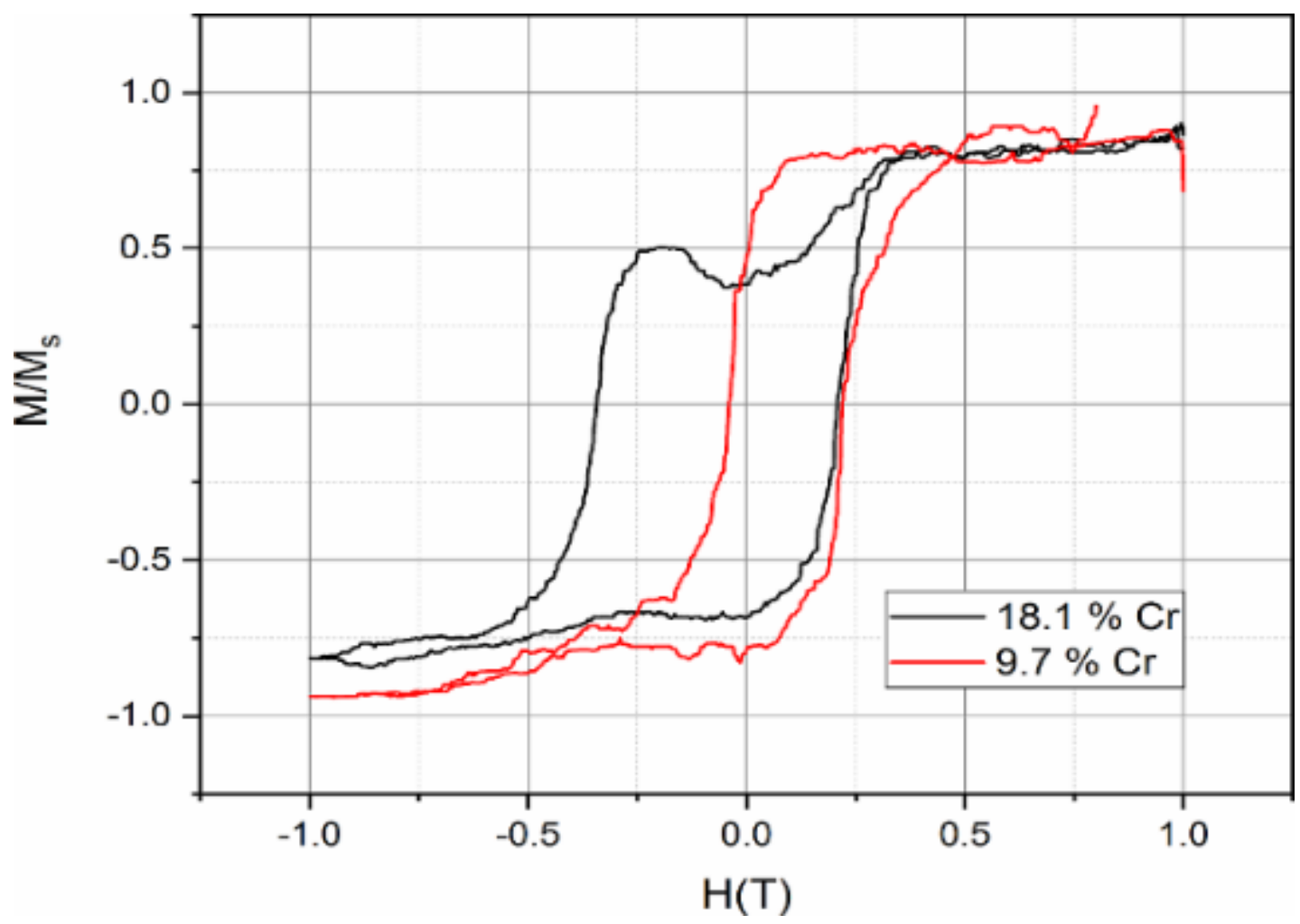

**Figure S18.** Hysteresis loops from spin dynamics simulations based on DFT-calculated parameters for cases of 18.1 (black) and 9.7 at. % (red) Cr concentration with respect to total core atoms, in core-shell NPs with Cr segregated at the surface of the core. The total diameter, *d*, of the NPs was 8 nm and the core-to-total diameter ratio, *c*, 0.2.

## Weight-Averaged Hysteresis Loops

To test how the behaviour of a macroscopic system consisting of non-agglomerating NPs can be simulated, simple weight-averaged hysteresis loops were estimated for the considered systems. The weights for each diameter were calculated assuming a log-normal distribution [S15,S16] with $\mu = 2.5$ and $\sigma = 0.2$, which yields a maximum at $d \sim 11$ nm, and subsequently normalised so that their sum was equal to 1. Cases for each core-to-total diameter ratio, *c*, were weighted equally. Two scenarios were considered. In the first, the averaging took place for the full range of *c* values ($c = 0.2$ to 0.7), while in the second only the cases with $c = 0.2$ to 0.5 were considered. The resulting loops and their characteristics are presented in Fig. S19 and Table S5. By averaging over the full range of the considered core sizes (0.2 to 0.7), the resulting values of $H_E$ and $H_C$ dropped by two orders of magnitude compared with their respective values for single NPs, due to the fact that low values of coercivity and exchange bias were found for NPs with large cores ($c \geq 0.5$). By averaging only in the range of $c = 0.2$ to 0.5, the resulting values of $H_E$ and $H_C$ dropped by only one order of magnitude, compared with the respective values for the single NPs. In both cases, the addition of Cr is again beneficial for $H_C$ and detrimental for $H_E$, but for the case of a homogeneous NiCr core (NiCr@NiO), the reduction of $H_E$ is smaller. Therefore, according to this description, the addition of Cr is beneficial for $H_E$ for NPs with well-defined diameter and core size.

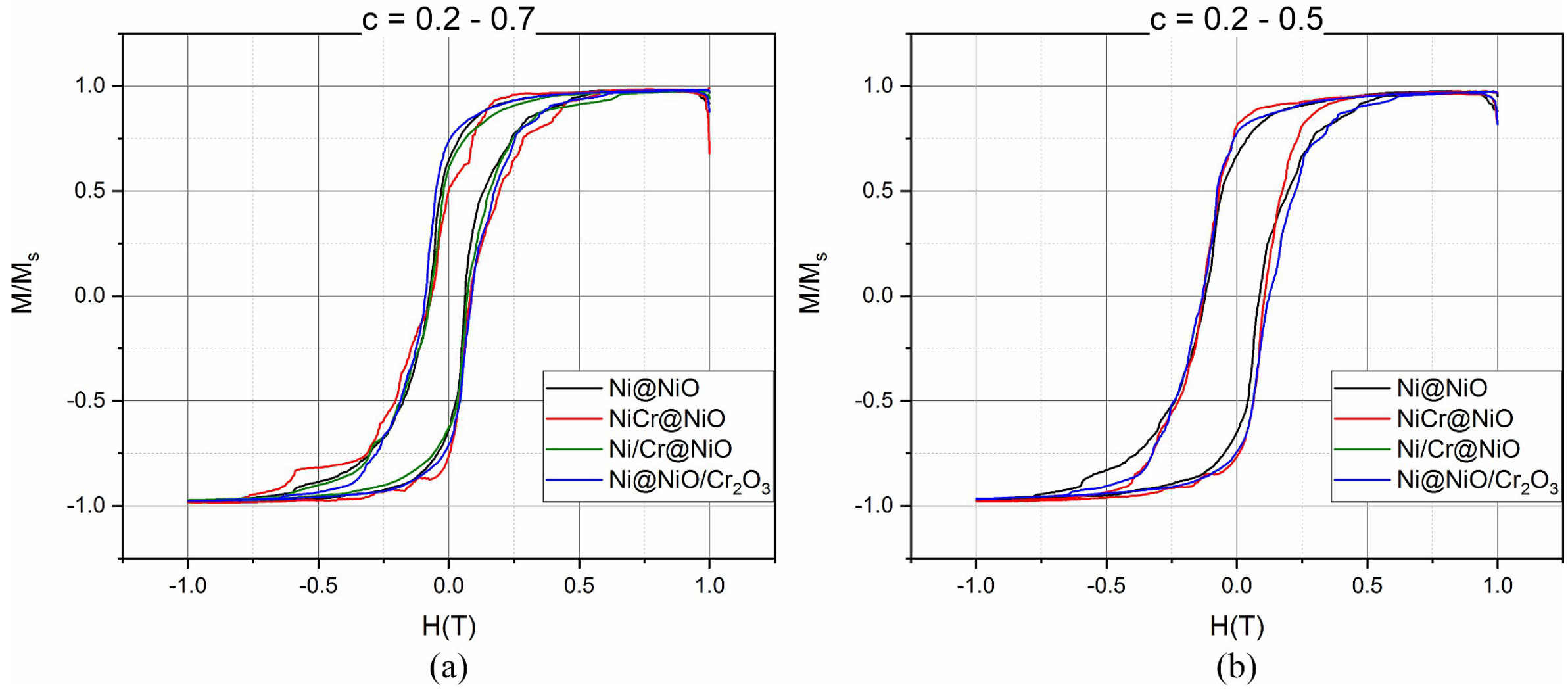

**Figure S19.** Averaged hysteresis loops for **a)** $c = 0.2$ - 0.7 and **b)** $c = 0.2$ - 0.4, for four considered NP configurations.

**Table S5.** Exchange bias field, $H_E$, and coercive field, $H_C$, for the cases considered in Figure S19.

| | $c$ = 0.2 - 07 | | $c$ = 0.2 - 0.5 | |
|---|---|---|---|---|
| | $H_E$ (kOe) | $H_C$ (kOe) | $H_E$ (kOe) | $H_C$ (kOe) |
| Ni@NiO | -0.10 | 0.73 | -0.18 | 1.16 |
| NiCr@NiO | -0.04 | 0.88 | -0.10 | 1.24 |
| Ni/Cr@NiO | -0.01 | 0.80 | -0.08 | 1.41 |
| Ni@NiO/$Cr_2O_3$ | -0.01 | 0.93 | -0.08 | 1.36 |

**Weight-Averaged Hysteresis Loops: Derivation of Nanoparticle Populations that Correspond to Experimental Samples**

This section constitutes documentation on how the different NP populations were calculated for the spin dynamics simulated magnetisation study to match the experimental depositions to as high a degree as possible.

In our spin dynamics investigation we examine 6 NP diameters, $d$ (nm):

- $d = 8$
- $d = 10$
- $d = 12$
- $d = 14$
- $d = 16$
- $d = 18$

These correspond to 6 NP volumes ($4/3\pi r^3$, where $r = d/2$, and $4/3\pi$ can be neglected as a constant):

- $V_8$ : 64 (elementary NP volume, $V$; units not required, as we are just comparing)
- $V_{10}$ : 125 ( $1.923 \times V$)
- $V_{12}$ : 216 ( $3.375 \times V$)
- $V_{14}$ : 343 ( $5.359 \times V$)
- $V_{16}$ : 512 ( $8 \times V$)
- $V_{18}$ : 729 ($11.391 \times V$)

Four main configurations and two variations were considered, corresponding to different chemical orderings (as shown in Figure 4a-e from the main article and Figure S16 from the Supporting Information).

- A : Ni@NiO
- $B_{FM}$ : NiCr@NiO
- $B_{PM}$: NiCr@NiO
- C : Ni/Cr@NiO
- D : Ni@NiO/$Cr_2O_3$
- $D_{\times 2}$: Ni@NiO/($Cr_2O_3 \times 2$)

Also, another four configurations were considered, corresponding to the elongated NPs of Figures S12-15.

Finally, 6 core-to-NP diameter ratios, $c$, were considered:

- $c = 0.2$
- $c = 0.3$
- $c = 0.4$
- $c = 0.5$
- $c = 0.6$
- $c = 0.7$

According to ref. [S17], the size distribution of NiCr NPs grown by magnetron-sputtering inert-gas condensation identical to our current deposition roughly follows a log-normal distribution, indicating partial NP coalescence (with hints of a bimodal, possibly due to coalesced NPs producing a minor second peak). It is reasonable to assume that the NPs follow the post-deposition size distribution of Figure S20, as presented in ref. [S17] following identical deposition conditions. However, the present study follows NPs after thermal treatment; therefore, present NPs likely appear larger due to (additional) oxidation upon thermal treatment.

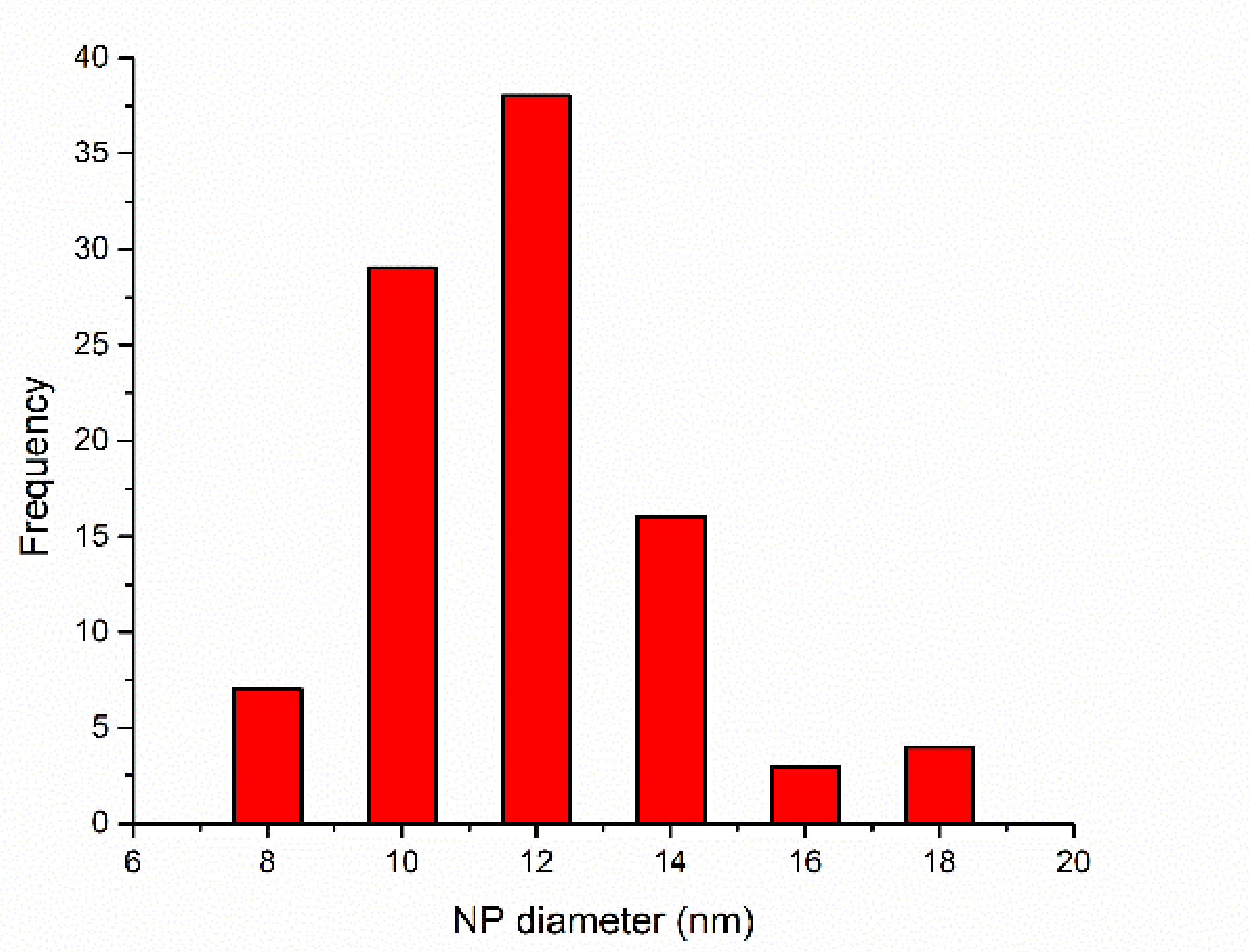

**Figure S20.** Post-deposition NP size distribution, based on scanning transmission electron microscopy analysis from ref. [S17]. The size distribution of our current samples is offset toward higher values due to annealing-assisted post-deposition oxidation.

This means, a representative sample may contain the following populations:

- 7 NPs : 8 nm
- 29 NPs : 10 nm
- 38 NPs : 12 nm
- 16 NPs : 14 nm
- 3 NPs : 16 nm
- 4 NPs : 18 nm

In total, 97 NPs.

The Cr content of each one of the NPs was calculated as:

1. **Ni@NiO (configuration A)**

0 at. % Cr at all cases.

2. **NiCr@NiO (configuration $B_{FM}$)**
3. **NiCr@NiO (configuration $B_{PM}$ )**
4. **Ni/Cr@NiO (configuration C)**

(identical Cr content but different configurations and corresponding magnetisation loops):

| | **d08** | **d10** | **d12** | **d14** | **d16** | **d18** |
|---|---|---|---|---|---|---|
| **c02** | 0.16% | 0.14% | 0.12% | 0.10% | 0.12% | 0.11% |
| **c03** | 0.63% | 0.37% | 0.36% | 0.35% | 0.35% | 0.45% |
| **c04** | 1.36% | 0.72% | 0.84% | 0.89% | 0.73% | 0.79% |
| **c05** | 1.99% | 1.74% | 1.56% | 1.46% | 1.35% | 1.52% |
| **c06** | 3.58% | 2.69% | 2.68% | 2.64% | 2.62% | 2.61% |
| **c07** | 5.96% | 4.89% | 4.95% | 5.04% | 5.06% | 5.10% |

5. **Ni@NiO/$Cr_2O_3$ (configuration D)**

| | **d08** | **d10** | **d12** | **d14** | **d16** | **d18** |
|---|---|---|---|---|---|---|
| **c02** | 12.43% | 12.30% | 12.43% | 12.49% | 12.34% | 12.46% |
| **c03** | 12.43% | 12.30% | 12.43% | 12.49% | 12.34% | 12.46% |
| **c04** | 12.43% | 12.30% | 12.43% | 12.49% | 12.34% | 12.46% |
| **c05** | 12.43% | 12.30% | 12.43% | 12.49% | 12.34% | 12.46% |
| **c06** | 12.43% | 12.30% | 12.43% | 12.49% | 12.34% | 12.46% |
| **c07** | 12.43% | 12.30% | 12.43% | 12.49% | 12.34% | 12.46% |

6. **$Ni@NiO/(Cr_2O_3 \times 2)$ (configuration $D_{\times 2}$)**

| | **d08** | **d10** | **d12** | **d14** | **d16** | **d18** |
|---|---|---|---|---|---|---|
| **c02** | 25.33% | 25.30% | 25.18% | 25.12% | 25.19% | 25.14% |
| **c03** | 25.33% | 25.30% | 25.18% | 25.12% | 25.19% | 25.14% |
| **c04** | 25.33% | 25.30% | 25.18% | 25.12% | 25.19% | 25.14% |
| **c05** | 25.33% | 25.30% | 25.18% | 25.12% | 25.19% | 25.14% |
| **c06** | 25.33% | 25.30% | 25.18% | 25.12% | 25.19% | 25.14% |
| **c07** | 25.33% | 25.30% | 25.18% | 25.12% | 25.19% | 25.14% |

We assigned Cr concentrations to each of these NPs, so that they correspond to the Cr content and configurations of the experimental samples as much as possible. We deposited 4 experimental samples with:

- 0
- 5
- 10
- 15 at. % Cr.

It was necessary to make educated assumptions in order to represent our experimental samples. For example, larger NPs should have higher *c* values, as they are less prone to oxidation (small NPs often oxidise completely). Also, according to ref. [S18], low Cr content allows Cr surface segregation to the surface of either the core or the shell. In contrast, Cr-rich NPs (>12 at. % Cr) should have almost all Cr diluted at the core. In that case, the cores should be PM due to the high Cr content. Of course, according to our configurations, this assumption would never allow for the Cr concentration of sample #4 to reach the nominal Cr concentration of 15 at. %. Therefore, we had to assume that half of the NPs would have a Cr surplus at the shell; such an assumption makes sense since it is possible for monometallic NPs to form during gas-phase synthesis, only to coalesce with each other at later stages of the nucleation and growth process. The probability of this happening increases with Cr nominal concentration, as it is more likely that pure-Cr nascent cluster form when the density of Cr atoms in the chamber is high.

The probabilities for each NP to have a specific structure was then estimated as:

| | A | B | C | D |
|---|---|---|---|---|
| **0 at. % Cr:** | 100% Ni@NiO | | | |
| **5 at. % Cr:** | | 20% NiCr@NiO $_{(FM)}$ | 40% Ni/Cr@NiO | 40% Ni@NiO/$Cr_2O_3$ |
| **10 at. % Cr:** | | 33% NiCr@NiO $_{(FM)}$ | | 67% Ni@NiO/$Cr_2O_3$ |
| **15 at. % Cr:** | | 50% NiCr@NiO $_{(PM)}$ | | 50% Ni@NiO/($Cr_2O_3$×2) |

For sample #1 (0 at. % Cr) the population of simulated NPs consisted of Ni@NiO in its entirety. According to the assumptions above, small NPs (*d*=8-12 nm) had a low core-to-NP diameter (*c*=0.2), whereas large NPs (*d*=14-18 nm) had a high core-to-NP diameter (*c*=0.7). Though not intentional, this choice also reflects the mixture of soft- and hard-magnetic exchanged coupled experimental C/S-NPs responsible for the "hummingbird"-type anisotropic magnetisation loops. Small NPs being more likely to deform, according to discussion in Fig. 5, also included a minority of oblong shapes.

According to a similar rationale, sample #2 (5 at. % Cr) consisted of a minority (20%) of NPs where Cr was still diluted in the FM core. Most of Cr atoms were surface-segregated. Since there is no way to distinguish the probability of them segregating to the core or the shell surface (and since both entities were detected experimentally), we divided them equally. A similar division between "small" and "large" NPs as in sample #1 would give too high a Cr concentration (6.33%); therefore, we used all *c* and *d* values in a way that small NPs contained small cores, and increasingly large NPs contained increasingly large cores (i.e., for *d*=8, *c*=0.2, for *d*=10, *c*=0.3, etc.). This treatment reduced the Cr content to 5.85%, closer to the experimental value.

For sample #3 (10 at. % Cr), the same "small" vs. "large" distinction was used as in sample #1; only this time ~33% of the NPs for each *c* and *d* value were considered NiCr@NiO$_{(FM)}$ and ~67% of the NPs were considered Ni@NiO/$Cr_2O_3$. This decision is meant to reflect the tendency of Cr atoms to surface-segregate. However, 10 at. % Cr is close enough to the limit where Cr atoms are trapped in the FM core, so a minority of NPs (33%) should reflect that fact.

As discussed above, sample #4 should mainly consist of NiCr@NiO(PM), since neighbouring Cr atoms need to overcome a significantly higher diffusion barrier in order to segregate to the surface.[S18] Due to the stochastic nature of gas-phase synthesis, however, the surplus of Cr atoms could help form segregated Cr-oxide islands, further assisted by the annealing-induced oxidation. As a result, and in order to increase the content of Cr atoms in the sample to levels comparable to the experimental concentration (13.99%), half of the NPs were of the NiCr@NiO(PM) configuration and the other half of the Ni@NiO/($Cr_2O_3$×2). All $c$ values were equally populated.

Naturally, minor adjustments were made to fit the actual numbers of atoms so that their sum was 97. The actual NP populations for each sample can be found in the spreadsheets also submitted as **Supporting Information**. The spreadsheets are facile tools that allow easy manipulation of the population numbers; furthermore, the interested reader may produce their own loops and calculate their numerical averages.

The overall Cr content, $X_{tot}$, of each sample can be calculated as follows:

$$X_{tot} = \frac{\sum_{i=1}^{97}(V_{NP}^{i} \times X_{NP}^{i})}{\sum_{i=1}^{97} V_{NP}^{i}} \ \% \quad \text{(S2)}$$

Let us see a couple of examples:

1. Sample consisting of two NPs:
    - NP1: volume $V$=1 and 10 at. % Cr => if NP1 has 100 atoms, 10 are Cr, 90 other
    - NP2: volume $V$=2 and 5 at. % Cr => if NP2 has 200 atoms, 10 are Cr, 190 other

In total, there are 300 atoms, 20 of which are Cr. Then, the at. % Cr is 20/300 × 100 % = 6.6 %

[(1 × 10) + (2 × 5)] / (1+2) % = 20 / 3 = 6.6

One needs to divide with the total number of atoms in the sample.

2. Imagine a sample consisting of three NPs:
    - NP1: volume $V$=1 and 10 at. % Cr => if NP1 has 100 atoms, 10 are Cr, 90 other
    - NP2: volume $V$=2 and 5 at. % Cr => if NP2 has 200 atoms, 10 are Cr, 190 other
    - NP3: volume $V$=5 and 5 at. % Cr => if NP3 has 500 atoms, 25 are Cr, 475 other

In total, there are 800 atoms, 45 of which are Cr. Then, the at. % Cr is 45/800 × 100 % = 5.625 %

[(1 × 10) + (2 × 5) + (5 × 5)] / (1 + 2 + 5) = 45 / 8 = 5.625